\documentclass[amsmath,amssymb,aps,prfluids,superscriptaddress,
]{revtex4-2}
\usepackage{multirow}
\usepackage{graphicx,dcolumn,bm,hyperref}
\usepackage[mathlines]{lineno}
\usepackage{amsmath,amsfonts, color,amssymb,subcaption,adjustbox,caption,booktabs}
\usepackage{placeins} 

\newcommand{\btau}{\boldsymbol{\tau}}

\newcommand{\bg}{{\bf g}}
\newcommand{\bE}{{\bf E}}
\newcommand{\bM}{{\bf M}}
\newcommand{\rM}{\rm M}

\newcommand{\re}{\rm e}
\newcommand{\bK}{{\bf K}}
\newcommand{\rK}{\rm K}
\newcommand{\bC}{{\bf C}}
\newcommand{\rC}{\rm C}

\newcommand{\bQ}{{\bf Q}}
\newcommand{\rQ}{\rm Q}

\newcommand{\bF}{{\bf F}}
\newcommand{\rrF}{\rm F}

\newcommand{\bv}{{\bm{v}}}

\newcommand{\bU}{{\bm{U}}}

\newcommand{\bu}{{\bm{u}}}
\newcommand{\bff}{{\bf f}}
\newcommand{\rff}{\rm f}
\newcommand{\rS}{\rm S}
\newcommand{\bx}{{\bf x}}
\newcommand{\rx}{\rm x}
\newcommand{\bxz}{{\bf x}_0}
\newcommand{\bxp}{{\bf x}_p}

\newcommand{\by}{{\bf y}}

\newcommand{\br}{{\bf r}}
\newcommand{\bn}{{\bf n}}

\newcommand{\bI}{{\bf I}}

\newcommand{\bzero}{{\bm{0}}}
\newcommand{\bsigma}{{\bm \sigma}}
\newcommand{\bt}{{\bf{t}}}
\newcommand{\bfu}{{\bf{f}}}

\newcommand{\bS}{{\bf{S}}}

\newcommand{\xhat}{{\bf{ \hat x}}}
\newcommand{\yhat}{{\bf \hat y}}
\newcommand{\zhat}{{\bf \hat z}}
\newcommand{\rhat}{{\bf \hat r}}
\newcommand{\phat}{{\bf \hat p}}

\newcommand{\Rdrop}{R_0}
\newcommand{\drop}{d}

\newcommand{\bnab}{{\bm{\nabla}}}

\newcommand{\visc}{\mu}

\newcommand{\viscratio}{\lambda}

\newcommand{\out}{+}
\newcommand{\ins}{-}
\newcommand{\surf}{S}

\newcommand{\st}{\gamma}

\newcommand{\refeqn}[1]{Eq. (\ref{#1})}

\newcommand{\refsec}[1]{Sec.~\ref{#1}}

\begin{document}

\preprint{}

\title{Translation of a spherical viscous drop driven by localized forcing in Stokes flow
}

\author{Sho Kawakami}
\author{Yuan-Nan Young}
\email{Corresponding author: yyoung@njit.edu}
\affiliation{Department of Mathematical Sciences,
New Jersey Institute of Technology, Newark, New Jersey 07102, USA}
\author{Howard A. Stone}
\email{Corresponding author: hastone@princeton.edu}
\affiliation{Department of Mechanical and Aerospace Engineering, Princeton University, Princeton, New Jersey 08544, USA}

\date{\today}

\begin{abstract}
Localized forcing in the fluid inside or outside a viscous drop can drive drop translation. Using the Lorentz reciprocal theorem, we derive an integral expression for the translational velocity of a spherical Newtonian drop subject to localized force and source distributions in either fluid and to interfacial traction. For a clean drop, we obtain explicit responses to Stokeslets, force dipoles, rotlets, general second force moments, and source dipoles as functions of position, orientation, and viscosity ratio. Interior forcing obeys a finite selection rule: only force moments through second order and the first source moment contribute directly to translation. Exterior forcing can couple to multipoles of all orders and produces distance-dependent responses. Although different enclosed singularities can produce the same drop velocity, resolving their exterior flows in drop-centered spherical Stokes modes provides additional constraints on the underlying forcing.  We also distinguish regularized force distributions, governed by prescribed kernel moments, from resolved rigid particles, governed by low-order surface-traction moments and prescribed slip. The framework unifies these representations and shows how exterior-flow measurements provide information beyond drop translation, laying the foundation for constructing squirmer-like viscous drop solutions with controllable far-field behaviors.

\end{abstract}

\maketitle

\section{Introduction}
The motion of drops, vesicles, and related deformable capsules driven by localized activity arises in applications ranging from particle-assisted transport and targeted delivery \cite{li2017micro,Lee:2023,malik2025magnetically} to synthetic cell-like systems powered by colloidal or biological components \cite{kokot2022spontaneous,Ramos:2020,rajabi2021directional}. Hydrodynamic interactions couple the motion of an enclosed or nearby particle to that of the interface. Predicting the resulting translational motion requires identifying which features of the localized forcing survive this hydrodynamic coupling.

An enclosed particle is commonly represented in one of two ways: by resolving its finite surface and imposing a traction or slip boundary condition, or by replacing it with a point force or a higher-order Stokes singularity.
The first retains particle geometry and near-field interactions, whereas the second gives a compact description of eccentric configurations. These representations need not be equivalent at finite size, even when they have the same point-particle limit.

Finite-size particles in confinement have been studied primarily in rigid enclosures \cite{zia:2016,chamolly2020stokes,marshall2021hydrodynamics,yariv2025mobility}. For fluid drops, concentric squirmers were considered in~\cite{reighPRF,reigh2017swimming}, followed by analyses of eccentric active particles in clean \cite{kree2021controlled,kree2022mobilities} and surfactant-covered drops \cite{shaik2018locomotion}. These studies establish the importance of particle position, activity, viscosity contrast, and interfacial rheology in the coupled particle--drop dynamics.

Point-singularity models sacrifice explicit particle-size information but permit analytical treatment. Stokeslets, force dipoles, and rotlets inside spherical drops or membranes have been examined in~\cite{daddi2018creeping,hoell2019creeping,sprenger2020towards,kree2021dynamics}, with extensions to more general singularities and weakly deformed clean or surfactant-laden drops \cite{kawakami2025migration}. Related models have been developed for spherical or nearly spherical confinement in Hele--Shaw geometries \cite{entov1993hele,tsang2015circularly,kawakami2025microswimmer}. Singularities outside a drop introduce a complementary problem because the translational response samples the spatially varying exterior mobility \cite{shaik2017point}.

Interfacial properties can qualitatively modify these responses. Tangential viscous tractions are continuous across a clean interface, whereas surfactant or interfacial elasticity generates tension gradients and Marangoni stresses. In the limit of a nearly inextensible drop interface, these stresses can suppress the motion produced by an enclosed force-free particle \cite{sprenger2020towards,kawakami2025migration}. The drop velocity therefore depends on both the localized forcing and interfacial stress transmission.

The Lorentz reciprocal theorem provides a direct route to global observables without first determining the complete Stokes flow \cite{masoud2019reciprocal}. Unlike approaches that construct the complete flow for selected
singularities, the reciprocal formulation gives the translational response
of arbitrary force and source distributions directly and exposes the
associated finite-moment selection rules. Among many applications, it has been used to calculate the migration of surfactant-covered drops in imposed flows \cite{subramanian1985stokes,pak2014viscous}. Here we derive a single reciprocal identity for the translation of a spherical Newtonian drop subject to arbitrary force and source distributions in either fluid and to a general interfacial traction. For a clean drop, the identity yields explicit velocities for interior and exterior Stokeslets, force dipoles, rotlets, second force moments, and source dipoles. The polynomial interior auxiliary field produces a finite moment selection rule, whereas the nonpolynomial exterior field admits contributions from multipoles of every order. 
We also obtain the far-field scaling of the drop velocity as a function of the position of the singularity for exterior singularities.


Drop velocity captures only the part of the forcing that drives translation; different singularities can therefore produce the same drop motion but different exterior flows.  We examine these exterior flows to distinguish such cases: we re-expand the flow of an enclosed singularity in drop-centered spherical Stokes modes and determine how these modes are transmitted to the exterior fluid.
The resulting eccentricity and far-field scalings show that singularities with the same translational projection can retain distinct exterior signatures, especially as the singularity approaches the interface. Finally, we extend the reciprocal formulation to regularized force distributions, finite rigid particles with distributed traction, active particles with surface slip, and nonuniform interfacial tension. This unified treatment separates the global drop response from the local representation of the activity and shows how exterior-flow measurements can further constrain the type, position, and orientation of localized interior forcing.

\section{Reciprocal formulation for localized forcing near a viscous drop}

\subsection{Setup and notation}
Consider the motion of a neutrally buoyant Newtonian spherical drop of radius $\Rdrop$ and viscosity $\viscratio\visc$ suspended in another Newtonian fluid of viscosity $\visc$ with interfacial tension $\st$ between the two fluids. 
The parameter $\viscratio$ is the viscosity ratio of the drop fluid to the suspending fluid.
The flow is driven by prescribed localized force or source distributions at an arbitrary position inside or outside the drop. 
These localized forcing and sources may result from the instantaneous hydrodynamic action of a particle, actuator, or other localized source of activity.
The Reynolds number is assumed sufficiently small so inertia is disregarded, and the capillary number $F/(\gamma\Rdrop)$ is also assumed sufficiently small such that the shape of the drop remains spherical.

\begin{figure}
\includegraphics[width=0.95\linewidth]{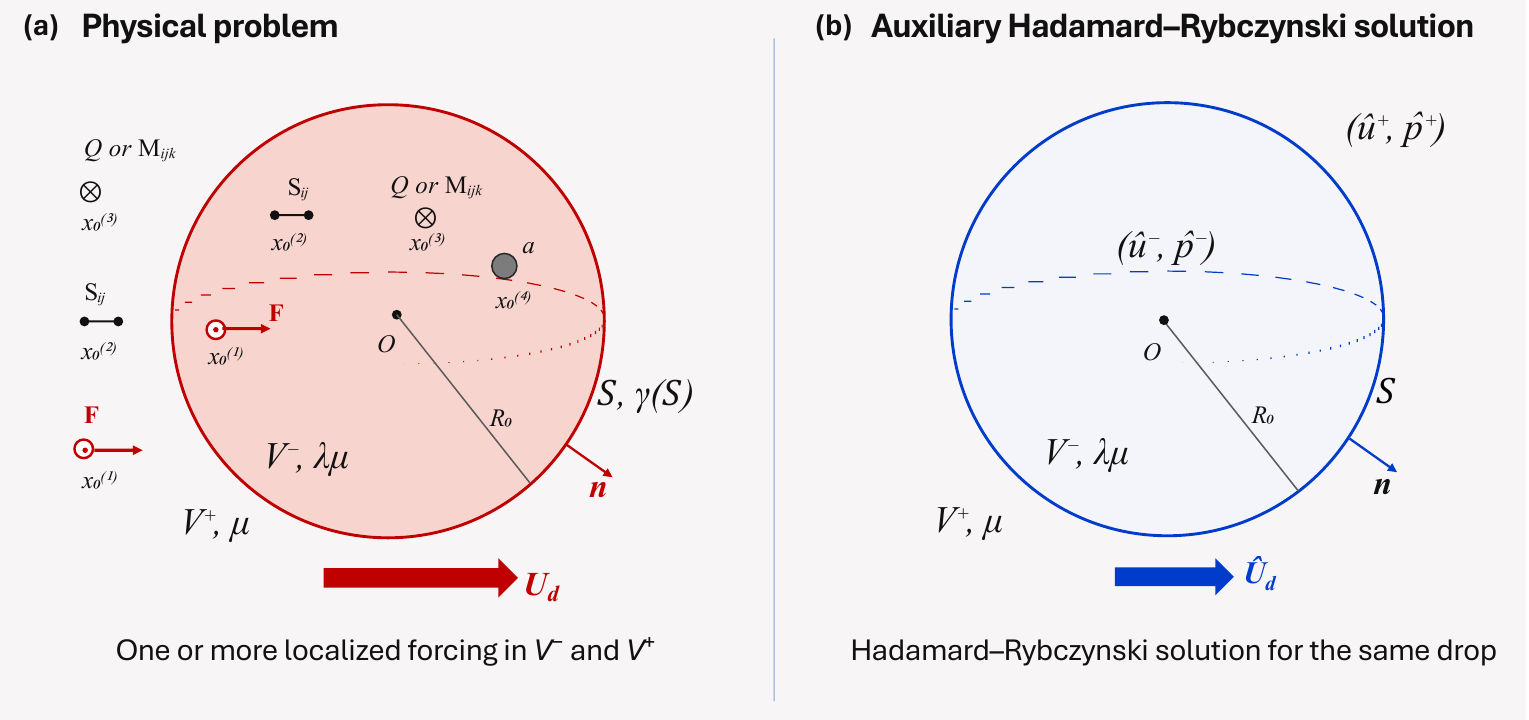}
    \caption{Setup of the two problems in the Lorentz reciprocal theorem. (a) Physical problem of a spherical viscous drop (of radius $R_0$) driven by different types of localized forcing at positions ${\bf x}_0^{(i)}$, with $i\in\{1,2,3\}$ (interior in Sec.~\ref{subsec:interior_singularities}, exterior in Sec.~\ref{subsec:exterior_singularities}) and active particle of finite size ($i=4$) in Sec.~\ref{sec:finite_size_effects}.  In addition, the drop tension $\gamma(\surf)$ may vary over the drop surface $\surf$, see Appendix \ref{sec:variable_tension}. (b) Model problem with an auxiliary (Hadamard-Rybczynski) solution for a viscous drop (with viscosity contrast $\lambda$) translating with velocity $\hat \bU_\drop$.}
    \label{fig:placeholder}
\end{figure}

We non-dimensionalize the lengths by the drop radius $\Rdrop$, the velocity by $F/(\visc \Rdrop)$, and the stress by $F/\Rdrop^2$. Note that $F$ is a reference force scale, and the singularity strengths
are nondimensionalized by the corresponding powers of $R_0$ (force dipole by $FR_0$, general second force moment by $F R_0^2$, and source dipole by $FR_0^2/\mu$).
All variables hereafter are dimensionless. 
Denoting 
the exterior fluid flow and pressure by $(\bu^\out,p^\out)$ and interior to the drop by $(\bu^\ins,p^\ins)$, the Newtonian stress tensors are
\begin{equation}
\bsigma^\out = -p^\out \bI + \big(\bnab \bu^\out + (\bnab \bu^{+})^T\big), \quad \bsigma^\ins = -p^\ins \bI + \lambda\big(\bnab \bu^\ins + (\bnab \bu^{-})^T\big).
\end{equation}
Both fluids are governed by the Stokes equations with sources:
\begin{subequations}
\label{GoverningEquations1}
\begin{align}
\text{exterior } (V^\out): & \quad \bnab \cdot \bsigma^\out  +\bff^\out(\bx,\bxz)= \bzero, \quad \bnab \cdot \bu^\out = s^\out(\bx,\bxz)  \\
\text{interior } (V^\ins): & \quad \bnab \cdot\bsigma^\ins  + \bfu^\ins(\bx,\bxz) = \bzero, \quad \bnab \cdot \bu^\ins =s^\ins(\bx,\bxz) .
\end{align}
\end{subequations}
$\bfu^\ins$ and $s^\ins$ correspond to vector and scalar point distributions modeling, respectively, the forces and sources (representing active particles) that drive fluid flow inside the drop. Likewise $\bfu^\out,s^\out$ are point distributions modeling the forces and sources driving fluid motion outside the drop. In this work we focus on sources such that $\int_{V^\ins}s^\ins dV = \int_{V^\out}s^\out dV = 0$.
An example of $\bff^\ins$ is a point force $\bF$  at $\bxz\in V^\ins$:
\begin{equation}
\bfu^\ins(\bx,\bxz) = \bF \, \delta(\bx - \bxz), \quad |\bxz| < 1,
\end{equation}
corresponding to a Stokeslet, the leading-order contribution of a forced particle, inside the drop,
together with $s^\ins=0, s^\out=0$,  and $\bff^\out = \bzero$.
On the interface $\surf$ (the unit sphere), we assume continuity of velocity and the traction jump in balance with surface traction $\bt_\surf$ generated only by surface tension and Marangoni stress in Eq.~\eqref{eqRTC:BC2}:
\begin{subequations}
\begin{align}
\bu^\out &= \bu^\ins, \qquad \bn \cdot \bsigma^\out - \bn \cdot \bsigma^\ins = \bt_\surf,\quad\text{on } \surf \label{eqRTC:BC1}\\
\bt_\surf &=\gamma(\bnab_\surf\cdot\bn)\bn-\bnab_\surf\gamma,\quad\text{on } \surf  \label{eqRTC:BC2}
\end{align}
\end{subequations}
where $\bn$ is the unit normal pointing from the interior to the exterior (Fig.~\ref{fig:placeholder}), and $\gamma$ is the surface tension that we assume to be constant ($\bnab_\surf\gamma=0$) except in Sec.~\ref{sec:variable_tension}. 


\subsection{Lorentz reciprocal theorem with interior and exterior body forces}

Let $(\bu, \bsigma, \bfu,p,s)$ and $(\hat{\bu}, \hat{\bsigma}, \hat{\bfu},\hat{p},\hat{s})$ be two Stokes flow fields in the same domain $V$ with boundary $S$, possibly with body forces and sources. The Lorentz reciprocal theorem corresponding to Eq.~(\ref{GoverningEquations1}) is
\begin{equation}
\int_{\surf} \bn \cdot \bsigma \cdot \hat{\bu} \, dS - \int_{\surf} \bn \cdot \hat{\bsigma} \cdot \bu \, dS = \int_V \left(\bu \cdot \hat{\bfu}-\hat{\bu} \cdot \bfu+\hat{p}\,s- p\,\hat{s}\right)\, dV.\label{eqRTC:RTgen}
\end{equation}
Unlike many typical applications of the reciprocal theorem, here the volume integral contributions from the forces $\bfu$ and sources $s$ are retained.

For the domain inside the drop, $V^\ins$, let $(\bu, \bsigma, \bfu,p,s) = (\bu^\ins, \bsigma^\ins, \bfu^\ins,p^\ins,s^\ins)$, corresponding to the problem we wish to solve, and $(\hat{\bu}, \hat{\bsigma},\hat{\bfu}, \hat{p},\hat{s}) = (\hat{\bu}^\ins, \hat{\bsigma}^\ins, \bzero,\hat{p}^\ins,0)$ in $V^\ins$ corresponding to an auxiliary problem that does not contain any forcing or source terms.
We denote the boundary by $S$ with outward normal $\bn$, and \refeqn{eqRTC:RTgen} gives
\begin{equation}
\int_{\surf} \bn \cdot \bsigma^\ins \cdot \hat{\bu}^\ins \, dS - \int_{\surf} \bn \cdot \hat{\bsigma}^\ins \cdot \bu^\ins \, dS = \int_{V^\ins}\left(\hat{p}^\ins\,s^\ins(\bx,\bxz) -\hat{\bu}^\ins \cdot \bfu^\ins(\bx,\bxz)\right) \, dV. \label{eqRTC:RTgeninner}
\end{equation}
The volume integrals contain information on the location and orientation of the singularity and their contributions to the flow.

Outside the drop, in $V^\out$, let $(\bu, \bsigma, \bfu,p,s) = (\bu^\out, \bsigma^\out, \bfu^\out,p^\out,s^\out)$ be the problem of interest and the auxiliary problem contain no forces or sources, $(\hat{\bu}, \hat{\bsigma},\hat{\bfu}, \hat{p},\hat{s}) = (\hat{\bu}^\out, \hat{\bsigma}^\out, \bzero,\hat{p}^\out,0)$.
The boundary $S^\out$ consists of $\surf$ and a sphere at infinity $\Gamma_\infty$, i.e., the limit of the integral over a sphere $S_a$ where the limit $a\to\infty$ is taken, with the outward normal at the interfaces given by $\bn^\out$ and $\bn^\infty$ respectively.
Then \refeqn{eqRTC:RTgen} gives
\begin{equation}
\begin{split}
&\int_{\surf} \bn^\out \cdot \bsigma^\out \cdot \hat{\bu}^\out \, dS - \int_{\surf} \bn^\out \cdot \hat{\bsigma}^\out \cdot \bu^\out \, dS +\int_{\Gamma_\infty} \bn^\infty \cdot \bsigma^\out \cdot \hat{\bu}^\out \, dS - \int_{\Gamma_\infty} \bn^\infty \cdot \hat{\bsigma}^\out \cdot \bu^\out \, dS\\
&= \int_{V^\out} \left (\hat{p}^\out\,s^\out(\bx,\bxz)-\hat{\bu}^\out \cdot \bfu^\out(\bx,\bxz) \right )\, dV. \label{eqRTC:RTgenouter}
\end{split}
\end{equation}
The original problem and the auxiliary problem (to be chosen later) will have the flow and stress fields that decay in the far field with rates at least $1/r$ and $1/r^2$ respectively. 
As a result, both integrals over $\Gamma_\infty$ vanish.
Since $\bn^\out = -\bn$ on $\surf$, \refeqn{eqRTC:RTgenouter} becomes
\begin{equation}
-\int_{\surf} \bn \cdot \bsigma^\out \cdot \hat{\bu}^\out \, dS + \int_{\surf} \bn \cdot \hat{\bsigma}^\out \cdot \bu^\out \, dS = \int_{V^\out} \left(\hat{p}^\out\,s^\out(\bx,\bxz)-\hat{\bu}^\out \cdot \bfu^\out(\bx,\bxz)\right) \, dV. \label{eqRTC:RTgenouter2}
\end{equation}
%
Adding \refeqn{eqRTC:RTgeninner} and \refeqn{eqRTC:RTgenouter2}, and using the continuity of velocity on $\surf$ ($\bu^\out = \bu^\ins =: \bu_\surf,$ $\hat{\bu}^\out = \hat{\bu}^\ins =: \hat{\bu}_\surf \text{ on } \surf$),
we obtain
\begin{equation}
\begin{split}
&\int_{\surf} \bn \cdot (\bsigma^\ins - \bsigma^\out) \cdot \hat{\bu}_\surf \, dS - \int_{\surf} \bn \cdot (\hat{\bsigma}^\ins - \hat{\bsigma}^\out) \cdot \bu_\surf \, dS \\
&\hspace{20pt}= \int_{V^\ins} \left(\hat{p}^\ins\,s^\ins(\bx,\bxz) -\hat{\bu}^\ins \cdot \bfu^\ins(\bx,\bxz)\right)\, dV+ \int_{V^\out} \left(\hat{p}^\out\,s^\out(\bx,\bxz) -\hat{\bu}^\out \cdot \bfu^\out(\bx,\bxz) \right)\, dV. \end{split}\label{eqRTC:RTgen2}
\end{equation}
Using boundary condition \refeqn{eqRTC:BC2} for the original problem and boundary condition $\bn\cdot(\hat\bsigma^+-\hat\bsigma^-) = \hat{\bt}_\surf$ for the auxiliary problem, we can rewrite \refeqn{eqRTC:RTgen2} in the form
\begin{equation}
\begin{split}
&-\int_{\surf} \bt_\surf \cdot \hat{\bu}_\surf \, dS + \int_{\surf} \hat{\bt}_\surf \cdot \bu_\surf \, dS \\
&\hspace{20pt}= \int_{V^\ins} \left(\hat{p}^\ins\,s^\ins(\bx,\bxz) -\hat{\bu}^\ins \cdot \bfu^\ins(\bx,\bxz)\right)\, dV+ \int_{V^\out} \left(\hat{p}^\out\,s^\out(\bx,\bxz) -\hat{\bu}^\out \cdot \bfu^\out(\bx,\bxz)\right) \, dV.  \end{split}\label{eqRTC:RTgen3}
\end{equation}
The integral identity in \refeqn{eqRTC:RTgen3} holds for any arbitrary Stokes singularity and arbitrary force balance on the interface.  In the following we use \refeqn{eqRTC:RTgen3} to compute the velocity of a clean drop driven by enclosed distributions and finite-size active particles.


\section{Translational motion of a clean drop driven by Point singularities\label{sec:clean_drop}}

To compute the drop translation velocity $\bU_d$ induced by a nearby Stokes singularity, choose the auxiliary problem (with hatted variables) in \refeqn{eqRTC:RTgen3} to be the rigid translation of the same drop with velocity $\hat{\bU}_\drop$ in an otherwise quiescent fluid, which is known as the Hadamard–Rybczynski solution. Here $r=\lvert\bx\lvert$, $\bx=r{\hat {\mathbf r}}$, and ${\bf n}= {\hat {\mathbf r}}$ on the spherical interface. The necessary parts of the auxiliary solution are 
\begin{subequations}
\begin{flalign}
    \hat{\bu}^\ins &=\frac{1}{2(\lambda+1)}\hat{\bU}_\drop\cdot\bK^\ins(\bx), \quad \bK^\ins(\bx) = (2\lambda+3-2\bx\cdot\bx)\bI+\bx\bx\label{eqRTC:buhat2}\\
    \hat{p}^\ins &= -\frac{5\lambda}{\lambda+1}\hat{\bU}_\drop\cdot\bx \label{eqRTC:phat}\\
    \hat{\bu}^\out &=\frac{1}{4(\lambda+1)}\hat{\bU}_\drop\cdot\bK^\out(\bx), \quad \bK^\out(\bx) = \frac{3\viscratio+2}{r}\bigg(\bI+\frac{\bx\bx}{r^2}\bigg)+\frac{\viscratio}{r^3}\bigg(\bI-3\frac{\bx\bx}{r^2}\bigg)\label{eqRTC:buhat2out}\\
    \hat{p}^\out &= \frac{3\viscratio+2}{2(\lambda+1)}\hat{\bU}_\drop\cdot\frac{\bx}{r^3} .\label{eqRTC:phatout}
\end{flalign}
\end{subequations}
Evaluation of the auxiliary velocity and traction jump at the interface gives
\begin{subequations}
\begin{flalign}
    \hat{\bu}_\surf &= \hat{\bU}_\drop\cdot\rhat\rhat+\frac{2\lambda+1}{2(\lambda+1)}\hat{\bU}_\drop\cdot(\bI-\rhat\rhat)\label{eqRTC:bushat}\\
    \hat{\bt}_\surf  &= -\frac{3(3\lambda+2)}{2(\lambda+1)}\hat{\bU}_\drop\cdot\rhat\rhat    .\label{eqRTC:tshat}
\end{flalign}
\end{subequations}
Then substituting Eqs.~(\ref{eqRTC:buhat2})-(\ref{eqRTC:tshat})
   into \refeqn{eqRTC:RTgen3}, we find
\begin{flalign}
\begin{split}
    &-\hat{\bU}_\drop\cdot\int_{\surf} \bt_\surf \cdot \bigg[\rhat\rhat+\frac{2\lambda+1}{2(\lambda+1)}(\bI-\rhat\rhat) \bigg]\, dS -\frac{3(3\lambda+2)}{2(\lambda+1)} \hat{\bU}_\drop\cdot\int_{\surf} \rhat\rhat \cdot \bu_\surf \, dS \\
    &\hspace{40pt}= \frac{1}{2(\viscratio+1)}\hat{\bU}_\drop \cdot\bigg[\int_{V^\ins} \left(-10\viscratio\bx s^\ins(\bx,\bxz)-\bK^\ins(\bx)\cdot\bff^\ins(\bx,\bxz)\right)\,dV\\
    &\hspace{150pt}+\int_{V^\out}\left(\frac{(3\viscratio+2)}{r^3}\bx s^\out(\bx,\bxz)-\frac{1}{2}\bK^\out(\bx)\cdot\bff^\out(\bx,\bxz)\right)\,dV\bigg].
    \end{split}\label{eqRTC:RTgen4}
\end{flalign}
For a clean drop $\bt_\surf = \gamma(\bnab_\surf\cdot\bn)\bn$, with $\bn=\rhat$ for a spherical drop. 
The first integral on the left-hand side of \refeqn{eqRTC:RTgen4} vanishes since it is odd in $\bn$.
Along with removal of the $\hat{\bU}_\drop$ term from \refeqn{eqRTC:RTgen4}, the equation can be further reduced to
\begin{flalign}
\begin{split}
    -\frac{3(3\lambda+2)}{2(\lambda+1)} \int_{\surf} \rhat\rhat \cdot \bu_\surf \, dS &= \frac{1}{2(\viscratio+1)}\bigg[\int_{V^\ins}\left( -10\viscratio\bx s^\ins(\bx,\bxz)-\bK^\ins(\bx)\cdot\bff^\ins(\bx,\bxz)\right)\,dV\\
    &\hspace{90pt}+\int_{V^\out}\left(\frac{(3\viscratio+2)}{r^3}\bx s^\out(\bx,\bxz)-\frac{1}{2}\bK^\out(\bx)\cdot\bff^\out(\bx,\bxz)\right)\,dV\bigg].\end{split}&&\label{eqRTC:RTgen5}
\end{flalign}
The translation velocity of the drop is obtained with the identity $\bU_\drop = \frac{3}{4\pi}\int_\surf\rhat\rhat\cdot\bu_\surf dS$ \cite{kawakami2025migration} and solving for $\bU_\drop$:
\begin{flalign}
\begin{split}
    \bU_\drop
    &= \frac{1}{4\pi(3\viscratio+2)}\bigg[\int_{V^\ins} \left(10\viscratio\bx s^\ins(\bx,\bxz)+\bK^\ins(\bx)\cdot\bff^\ins(\bx,\bxz)\right)\,dV\\
    &\hspace{90pt}+\int_{V^\out}\left(-\frac{(3\viscratio+2)}{r^3}\bx\, s^\out(\bx,\bxz)+\frac{1}{2}\bK^\out(\bx)\cdot\bff^\out(\bx,\bxz)\right)\,dV\bigg].
    \end{split}\label{eqRTC:Udropeqn}
\end{flalign}
\refeqn{eqRTC:Udropeqn} gives the translational velocity of a clean drop driven by arbitrary force and source distributions. Section~\ref{subsec:interior_singularities} establishes the finite selection rule for interior point singularities, Sec.~\ref{subsec:exterior_singularities} treats exterior point singularities, and Sec.~\ref{subsec:interior_exterior_flow} examines the exterior-flow signatures of enclosed singularities.

\subsection{Translational response to interior singularities}
\label{subsec:interior_singularities}
Here we use the reciprocal identity in Eq.~\eqref{eqRTC:Udropeqn} to compute the drop translational velocity for various localized force and source distributions, and summarize the results in Table~\ref{table:singularitysummary}.

\subsubsection{Stokeslet}
\label{subsubsec:interiorStokeslet}

For a Stokeslet of strength $\bF$ at $\bxz$, with $|\bxz|<1$,
\begin{equation}
\bff^\ins=\bF\delta(\bx-\bxz),
\qquad
s^\ins=0,
\qquad
\bff^\out=\bzero,
\qquad
s^\out=0.
\end{equation}
Eq.~\eqref{eqRTC:Udropeqn} gives
\begin{equation}
\bU_\drop
=
\frac{1}{4\pi(3\lambda+2)}
\left[
(2\lambda+3-2|\bxz|^2)\bI+\bxz\bxz
\right]\cdot\bF,
\label{eqRTC:RTStokeslet1}
\end{equation}
in agreement with ~\cite{kawakami2025migration,kree2021controlled}.

\subsubsection{Force dipoles}
\label{subsubsec:interiorforcedipole}

For the force dipoles, the force at $\bxz$ is
\begin{equation}
\rff_i^\ins(\bx)=\rS_{ij}\,\partial_j\delta(\bx-\bxz),
\qquad
s^\ins=0,
\label{eq:force_dipole_definition}
\end{equation}
where $\bS$ is a rank-two tensor. The derivative of the interior kernel is
\begin{equation}
\partial_j\rK^\ins_{mi}
=
-4\rx_j\delta_{mi}+\delta_{mj}\rx_i+\rx_m\delta_{ij}.
\label{eq:gradKminus}
\end{equation}
After one integration by parts,
\begin{align}
U_{\drop,m}
&=
-\frac{\rS_{ij}\,\partial_j\rK^\ins_{mi}(\bxz)}
{4\pi(3\lambda+2)} =
\frac{
4\rS_{mj}\rx_{0j}-\rS_{im}\rx_{0i}-\rS_{ii}\rx_{0m}
}{4\pi(3\lambda+2)}.
\label{eq:drop_velocity_force_dipole}
\end{align}
Equivalently,
\begin{equation}
\bU_\drop
=
\frac{
4\bS\cdot\bxz-\bS^T\cdot\bxz
-\operatorname{tr}(\bS)\bxz
}{4\pi(3\lambda+2)}.
\label{eq:drop_velocity_force_dipole_vector}
\end{equation}
Thus an interior force dipole couples linearly to its displacement from the
drop center.

\noindent{Axisymmetric force dipole:}
For an axisymmetric force dipole $\bS=-\mathcal P\phat\phat$, where $\phat$ is a unit vector,
\begin{equation}
\bU_\drop
=
\frac{\mathcal P}{4\pi(3\lambda+2)}
\left[\bxz-3(\phat\cdot\bxz)\phat\right],
\label{eq:axisymmetric_force_dipole_velocity}
\end{equation}
in agreement with \cite{kawakami2025migration}. With this convention,
$\mathcal P>0$ and $\mathcal P<0$ correspond, respectively, to the pusher and
puller signs used in the squirmer literature.

\noindent{Rotlet:}
For a rotlet with  $\btau$ denoting the physical point torque, the body-force distribution is
\begin{equation}
\bff^\ins
=-\frac12\btau\times\bnab\delta(\bx-\bxz),
\qquad
\bS_{ij}=\frac12\varepsilon_{ijk}\tau_k.
\label{eq:rotlet_force_density}
\end{equation}
The factor $1/2$ ensures that
$\int(\bx-\bxz)\times\bff^\ins\,dV=\btau$. Eq.~\eqref{eq:drop_velocity_force_dipole}
then gives
\begin{equation}
\bU_\drop
=-\frac{5}{8\pi(3\lambda+2)}\btau\times\bxz,
\label{eq:interior_rotlet_velocity}
\end{equation}
which agrees with the physical-torque convention in \cite{kawakami2025migration}.

\subsubsection{General second force moment}
\label{subsubsec:interiorquadrupole}

A general second force moment is represented by
\begin{equation}
\rff_i^\ins(\bx)
=\rM_{ijk}\,\partial_j\partial_k\delta(\bx-\bxz),
\qquad \rM_{ijk}=\rM_{ikj}.
\label{eq:force_quadrupole_definition}
\end{equation}
The second derivative of the interior kernel is constant:
\begin{equation}
\partial_j\partial_k\rK^\ins_{mi}
=
-4\delta_{jk}\delta_{mi}
+\delta_{mj}\delta_{ik}
+\delta_{mk}\delta_{ij}.
\label{eq:gradgradKminus}
\end{equation}
Consequently,
\begin{equation}
U_{\drop,m}
=
\frac{\rM_{ijk}
\left(
-4\delta_{jk}\delta_{mi}
+\delta_{mj}\delta_{ik}
+\delta_{mk}\delta_{ij}
\right)}{4\pi(3\lambda+2)},
\label{eq:drop_velocity_force_quad}
\end{equation}
which implies that $\bU_\drop=\bzero$ for a fully trace-free irreducible quadrupole. Here the trace of a third-order tensor denotes a
contraction over two indices and is therefore a vector.
However, the axisymmetric second force moment
\begin{align}
\mathcal M_{ijk}=M_0\hat p_i\hat p_j\hat p_k,
\qquad |\hat{\mathbf p}|=1,
\end{align}
is not trace-free.    In particular, $\mathcal M_{ijj}
=
M_0\hat p_i,$ and $\mathcal M_{jji}
=
M_0\hat p_i$.
Hence the degree-one combination
\(P_i=2\mathcal M_{ijj}-\mathcal M_{jji}\) is
\(P_i=M_0\hat p_i\), and the resulting drop velocity is
\begin{equation}
\bU_\drop
=-\frac{M_0}{2\pi(3\lambda+2)}\phat.
\label{eq:axisymmetric_force_quadrupole_velocity}
\end{equation}
Unlike the dipole contribution, this explicit quadrupole projection is
independent of $\bxz$.

\subsubsection{Source dipole}
\label{subsubsec:interiorsourcedipole}

For the convention
\begin{equation}
\bff^\ins=\bzero,
\qquad
s^\ins=-\bQ\cdot\bnab\delta(\bx-\bxz),
\label{eq:source_dipole_definition}
\end{equation}
Eq.~\eqref{eqRTC:Udropeqn} gives
\begin{align}
\bU_\drop
&=
-\frac{5\lambda}{2\pi(3\lambda+2)}
\int_{V^\ins}
\bx\,[\bQ\cdot\bnab\delta(\bx-\bxz)]\,dV = \frac{5\lambda}{2\pi(3\lambda+2)}\bQ.
\label{eqRTC:RTSD3}
\end{align}
The result is independent of $\bxz$. Its sign is fixed by
Eq.~\eqref{eq:source_dipole_definition}; comparisons with other conventions
must transform the definition of $\bQ$ accordingly
\cite{kawakami2025migration}.

\subsubsection{Higher-order interior multipoles\label{subsubsec:interior_HO_multipoles}}

The selection rule for higher-order interior singularities follows directly from
$\partial_j\partial_k\partial_l\rK^\ins_{mi}=0,$
$\partial_j\partial_k \rx_m=0.$
Therefore, every force multipole containing three or more derivatives of the
delta distribution---a force octupole or higher---has zero direct projection
onto $\bU_\drop$. Likewise, source multipoles containing two or more
derivatives---a source quadrupole or higher---do not contribute directly.
These singularities may still alter the local flow and stress, but they are
orthogonal to the translating-drop auxiliary mode.

\begin{table}
\begin{center}
\begin{tabular}{|c|c|c c|}
\toprule

\textbf{Singularity type} & \textbf{Exact drop velocity $\bU_\drop$}&\multicolumn{2}{|c|}{\parbox{150pt}{\textbf{Example schematic representation}}} \\
\midrule
\parbox{130pt}{Point Force (Stokeslet)\\\refsec{subsubsec:interiorStokeslet}\\$\bfu=\bF\delta(\bx-\bxz)$} & $\bU_d=\frac{\big[(2\lambda+3-2|\bxz|^2)\bI+\bxz\bxz\big]\cdot\bF}{4\pi(3\lambda+2)} $ Eq.~\eqref{eqRTC:RTStokeslet1}&\parbox{100pt}{Stokeslet:}&\adjustimage{height=50pt,valign=m}{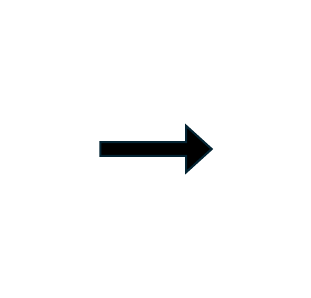} \\
\midrule
\multirow[c]{2}{*}{%
  \parbox[c]{130pt}{%
    \centering
    Force Dipole\\    \refsec{subsubsec:interiorforcedipole}\\
    $\rff_i=\rS_{ij}\bnab_j\delta(\bx-\bxz)$
  }%
}
&
\parbox[c]{185pt}{%
  \centering
  Force dipole:\\[2pt]
  $\displaystyle
  U_{d,m}
  =
  \frac{
    4\rS_{mj}\rx_{0j}
    -\rS_{im}\rx_{0i}
    -\rS_{ii}\rx_{0m}
  }{
    4\pi(3\lambda+2)
  }
  $
  Eq.~\eqref{eq:drop_velocity_force_dipole}
}
&
\parbox[c]{100pt}{%
  \centering
  Force dipole:
}
&
\adjustimage{height=50pt,valign=m}
  {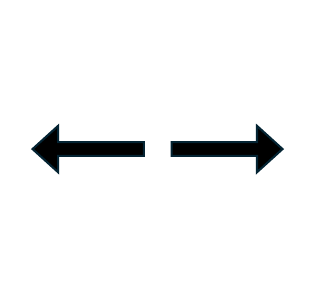}
\\
&
\parbox[c]{175pt}{%
  \centering
  Rotlet:\\[2pt]
  $\displaystyle
  \bU_\drop
  =
  -\frac{5}{8\pi(3\lambda+2)}
  \btau\times\bxz
  $
  Eq.~\eqref{eq:interior_rotlet_velocity}
}
&
\parbox[c]{100pt}{%
  \centering
  Rotlet:
}
&
\adjustimage{height=50pt,valign=m}
  {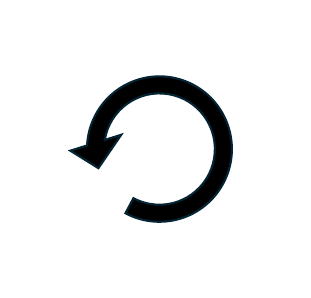}
\\
\midrule
\parbox{130pt}{General Second Force Moment \\\refsec{subsubsec:interiorquadrupole}\\$\rff_i=\rM_{ijk}\bnab_j\bnab_k\delta(\bx-\bxz)$} & $U_{d,m}=\frac{\rM_{ijk}(-4\delta_{jk}\delta_{mi}+\delta_{mj}\delta_{ik}+\delta_{mk}\delta_{ij})}{4\pi(3\lambda+2)}$ Eq.~\eqref{eq:drop_velocity_force_quad}&\parbox{100pt}{Axisymmetric\\ second force moment:}&\adjustimage{height=50pt,valign=m}{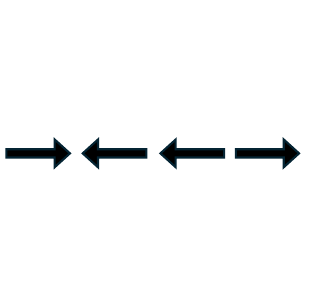} \\
\midrule
\parbox{130pt}{Source Dipole \\\refsec{subsubsec:interiorsourcedipole}\\$s=-\bQ\cdot\bnab\delta(\bx-\bxz)$} & $\bU_d=\frac{5\lambda}{2\pi(3\lambda+2)}\bQ$ Eq.~\eqref{eqRTC:RTSD3}&\parbox{100pt}{Source Dipole:}&\adjustimage{height=50pt,valign=m}{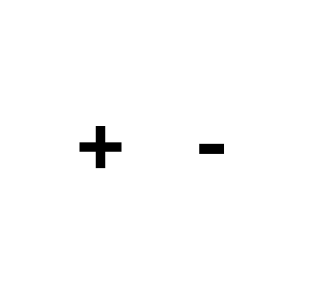}\\
\midrule
\parbox{120pt}{Higher-order force ($\cal F$) and source ($\cal Q$) multipoles\\
Sec.~\ref{subsubsec:interior_HO_multipoles}} &$\bzero$& & \\

\bottomrule
\end{tabular}
\end{center}
\caption{Summary of drop velocity induced by an enclosed singularity.} \label{table:singularitysummary}

\end{table}

\subsection{Translational response to exterior singularities}
\label{subsec:exterior_singularities}

Here we distinguish the exact exterior responses from the far-field summary in
Table~\ref{table:exterior_singularity_summary}.  The equations below are valid
for any $r_0:=|\bxz|>1$, whereas Table~\ref{table:exterior_singularity_summary}
retains only the dominant contribution for $r_0\gg1$.  Writing
$\mathbf e_0=\bxz/r_0$, the exterior auxiliary kernel in
Eq.~\eqref{eqRTC:buhat2out} contains an $O(r^{-1})$ part and an
$O(r^{-3})$ part.  The former determines the leading far-field response to an
exterior force multipole, while the latter gives a viscosity-dependent
correction two powers of $r_0$ smaller.  Exterior source multipoles instead
project against $\bx/r^3=O(r^{-2})$ in Eq.~\eqref{eqRTC:Udropeqn}.

\subsubsection{Stokeslet\label{subsubsec:Stokeslet}}

For a Stokeslet of strength $\bF$ at $\bxz$, with $|\bxz|>1$,
\begin{equation}
\bff^\out=\bF\delta(\bx-\bxz),
\qquad
\bff^\ins=\bzero,
\qquad
s^\ins=s^\out=0.
\end{equation}
The exterior kernel, not the interior kernel, must be used. Thus
\begin{align}
\bU_\drop
&=
\frac{1}{8\pi(3\lambda+2)}\bK^\out(\bxz)\cdot\bF =
\frac{1}{8\pi r_0}
\left(\bI+\mathbf e_0\mathbf e_0\right)\cdot\bF
+
\frac{\lambda}{8\pi(3\lambda+2)r_0^3}
\left(\bI-3\mathbf e_0\mathbf e_0\right)\cdot\bF.
\label{eq:exterior_stokeslet_velocity}
\end{align}
For $r_0\gg1$, the first term dominates,
\begin{equation}
\bU_\drop
\sim
\frac{1}{8\pi r_0}
\left(\bI+\mathbf e_0\mathbf e_0\right)\cdot\bF
=O(r_0^{-1}),
\label{eq:exterior_stokeslet_farfield}
\end{equation}
which is the Stokeslet entry in Table~\ref{table:exterior_singularity_summary}.
The viscosity-dependent term in Eq.~\eqref{eq:exterior_stokeslet_velocity} is
subdominant, $O(r_0^{-3})$.

\subsubsection{Force dipole\label{subsubsec:FD}}

Let
\begin{equation}
\rff_i^\out=\rS_{ij}\partial_j\delta(\bx-\bxz),
\qquad
\bff^\ins={\bf 0},
\qquad
s^\ins=s^\out=0.
\end{equation}
The derivative of the exterior kernel is
\begin{align}
\partial_j\rK^\out_{mi}
&=
\frac{3\lambda+2}{r^3}
\left(
\delta_{mj}\rx_i+\delta_{ij}\rx_m-\delta_{mi}\rx_j
-\frac{3\rx_m\rx_i\rx_j}{r^2}
\right)-
\frac{3\lambda}{r^5}
\left(
\delta_{mj}\rx_i+\delta_{ij}\rx_m+\delta_{mi}\rx_j
-\frac{5\rx_m\rx_i\rx_j}{r^2}
\right).
\label{eq:gradKplus}
\end{align}
Hence
\begin{equation}
U_{\drop,m}
=
-\frac{\rS_{ij}\partial_j\rK^\out_{mi}(\bxz)}
{8\pi(3\lambda+2)}.
\label{eq:exterior_force_dipole_velocity}
\end{equation}
For an exterior rotlet with physical point torque $\btau$, the
body-force distribution and corresponding antisymmetric dipole tensor are
$\bff^\out
=
-\frac{1}{2}\btau\times\bnab\delta(\bx-\bxz),$
$\rS_{ij}=\frac{1}{2}\epsilon_{ijk}\tau_k.$
The antisymmetry of $\rS_{ij}$ makes its contraction with the second
bracket in Eq.~\eqref{eq:gradKplus} vanish, while its contraction with
the first bracket reduces to $(\btau\times\bx)_m$.  Equation
\eqref{eq:exterior_force_dipole_velocity} therefore gives the exact
rotlet-induced drop velocity
\begin{equation}
\bU_\drop
=
-\frac{\btau\times\bxz}{8\pi r_0^3}
=
-\frac{\btau\times\re_0}{8\pi r_0^2},
\qquad r_0>1.
\label{eq:exterior_rotlet_velocity}
\end{equation}
Thus the exterior-rotlet response is independent of the viscosity ratio
$\lambda$.

Unlike Eq.~\eqref{eq:drop_velocity_force_dipole}, the general
force-dipole weight is nonpolynomial in $\bxz$.  The two terms in
Eq.~\eqref{eq:gradKplus} scale as $r^{-2}$ and $r^{-4}$, respectively.
Therefore
\begin{align}
U_{\drop,m}
&\sim
-\frac{\rS_{ij}}{8\pi r_0^2}
\Big(
\delta_{mj}\re_{0i}
+\delta_{ij}\re_{0m}
-\delta_{mi}\re_{0j}
-3\re_{0m}\re_{0i}\re_{0j}
\Big)
=O(r_0^{-2}).
\label{eq:exterior_force_dipole_farfield}
\end{align}
This is the force-dipole entry in
Table~\ref{table:exterior_singularity_summary}.  A symmetric force
dipole and an antisymmetric rotlet both have an $r_0^{-2}$ leading
decay, although their tensor contractions differ.  For the symmetric
part of a general force dipole, the viscosity-dependent correction
enters at $O(r_0^{-4})$ when its contraction is nonzero.  For the
rotlet, however, this correction cancels identically, and
Eq.~\eqref{eq:exterior_rotlet_velocity} is exact.
\subsubsection{General second force moment\label{subsubsec:exteriorquadrupole}}

For an exterior general second force moment,
\begin{equation}
\rff_i^\out
=\rM_{ijk}\partial_j\partial_k\delta(\bx-\bxz),
\qquad
\bff^\ins={\bf 0},
\qquad
s^\ins=s^\out=0,
\end{equation}
Eq.~\eqref{eqRTC:Udropeqn} gives
\begin{equation}
U_{\drop,m}
=
\frac{\rM_{ijk}}{8\pi(3\lambda+2)}
\left.
\partial_j\partial_k\rK^\out_{mi}(\bx)
\right|_{\bx=\bxz}.
\label{eq:exterior_force_quadrupole_velocity}
\end{equation}
The dominant contribution comes from the $O(r^{-1})$ part of
$\bK^\out$, so that
\begin{equation}
U_{\drop,m}
\sim
\frac{\rM_{ijk}}{8\pi}
\left.
\partial_j\partial_k
\left[
\frac{1}{r}
\left(
\delta_{mi}+\frac{x_mx_i}{r^2}
\right)
\right]
\right|_{\bx=\bxz}
=O(r_0^{-3}).
\label{eq:exterior_force_quadrupole_farfield}
\end{equation}
Thus the exterior general second-force-moment (force quadrupole) contribution decays as $r_0^{-3}$, as listed in
Table~\ref{table:exterior_singularity_summary}; the first
viscosity-dependent correction is $O(r_0^{-5})$.

\subsubsection{Source dipole\label{subsubsec:SD}}

For
\begin{equation}
\bff^\ins=\bff^\out=\bzero,
\qquad
s^\ins=0,
\qquad
s^\out=-\bQ\cdot\bnab\delta(\bx-\bxz),
\end{equation}
Eq.~\eqref{eqRTC:Udropeqn} gives
\begin{align}
\bU_\drop
&=
-\frac{1}{4\pi}
\bQ\cdot\bnab
\left(\frac{\bx}{r^3}\right)_{\bx=\bxz}
= -\frac{1}{4\pi r_0^3}
\left(
\bI-3\mathbf e_0\mathbf e_0
\right)\cdot\bQ.
\label{eq:exterior_source_dipole_velocity}
\end{align}
Here the $r_0^{-3}$ expression is exact, not only asymptotic, and therefore
coincides with the source-dipole entry in
Table~\ref{table:exterior_singularity_summary}.  In contrast to the exterior
force singularities above, this translational projection contains no
viscosity-ratio dependence.  

\subsubsection{Higher-order exterior multipoles\label{subsubsec:HO}}

The exterior kernel is nonpolynomial, so its derivatives do not terminate.
For a rank-$n$ force multipole written as
\begin{equation}
\rff_i^\out
={\cal F}_{i j_1\cdots j_{n-1}}
\partial_{j_1}\cdots\partial_{j_{n-1}}
\delta(\bx-\bxz),
\end{equation}
its contribution is
\begin{equation}
U_{\drop,m}
=
\frac{(-1)^{n-1}}{8\pi(3\lambda+2)}
{\cal F}_{i j_1\cdots j_{n-1}}
\left.
\partial_{j_1}\cdots\partial_{j_{n-1}}\rK^\out_{mi}(\bx)
\right|_{\bx=\bxz}.
\label{eq:exterior_force_multipole_general}
\end{equation}
Since the leading part of $\bK^\out$ is $O(r^{-1})$, a rank-$n$ force
multipole has
\begin{equation}
\bU_\drop=O(r_0^{-n}),
\qquad r_0\gg1,
\label{eq:exterior_force_multipole_scaling}
\end{equation}
with a viscosity-dependent correction of order $r_0^{-(n+2)}$.

To use the same sign convention as the source dipole above, write an
order-$n$ source multipole as
\begin{equation}
s^\out
=-{\cal Q}_{j_1\cdots j_n}
\partial_{j_1}\cdots\partial_{j_n}
\delta(\bx-\bxz).
\end{equation}
Then
\begin{equation}
U_{\drop,m}
=
\frac{(-1)^n}{4\pi}
{\cal Q}_{j_1\cdots j_n}
\left.
\partial_{j_1}\cdots\partial_{j_n}
\left(\frac{\rx_m}{r^3}\right)
\right|_{\bx=\bxz},
\label{eq:exterior_source_multipole_general}
\end{equation}
and hence
\begin{equation}
\bU_\drop=O(r_0^{-(n+2)}),
\qquad r_0\gg1.
\label{eq:exterior_source_multipole_scaling}
\end{equation}
Thus exterior multipoles of arbitrarily high order can contribute in
principle, although particular tensor symmetries may make an individual
contraction vanish.  Table~\ref{table:exterior_singularity_summary} is
therefore an asymptotic summary of the exact expressions above; for a
singularity close to the drop, $r_0=O(1)$, the full formulas rather than only
the leading terms should be used.

\begin{table}
\begin{center}
\small
\begin{tabular}{|c|c|c c|}
\toprule

\textbf{Singularity type} & \textbf{Leading far-field drop velocity $\bU_\drop,\;\; r_0 > 1$}&\multicolumn{2}{|c|}{\parbox{150pt}{\textbf{Example schematic representation}}} \\
\midrule
\parbox[c]{130pt}{%
\centering
Point Force (Stokeslet)\\
Sec.~\ref{subsubsec:Stokeslet}\\
$\mathbf f^{+}
 =\mathbf F\,\delta(\mathbf x-\mathbf x_0)$
}
&
\parbox[c]{185pt}{%
\centering
$\displaystyle
\mathbf U_d
\sim
\frac{1}{8\pi r_0}
\left(
\mathbf I+
{\mathbf e}_0{\mathbf e}_0
\right)\cdot\mathbf F
$
\\[2pt]
$\displaystyle
=O(r_0^{-1})
$ Eq.~\eqref{eq:exterior_stokeslet_farfield}
}
&
\parbox[c]{90pt}{%
\centering
Stokeslet:
}
&
\adjustimage{height=50pt,valign=m}
{IllustrationA.pdf}
\\
\midrule
\parbox[c]{130pt}{%
\centering
Force Dipole\\Sec.~\ref{subsubsec:FD}\\
$\rff_i^{+}
 =\rS_{ij}\,\partial_j
 \delta(\mathbf x-\mathbf x_0)$
}
&
\parbox[c]{185pt}{%
\centering
Force dipole:\\[2pt]
$\displaystyle
U_{d,m}
\sim
-\frac{\rS_{ij}}{8\pi r_0^2}
$
\\[-1pt]
$\displaystyle
\times
\Big(
\delta_{mj}\re_{0i}
+\delta_{ij}\re_{0m}
-\delta_{mi}\re_{0j}
-3\re_{0m}
 \re_{0i}
 \re_{0j}
\Big)
$
\\[-1pt]
$\displaystyle
=O(r_0^{-2})
$ Eq.~\eqref{eq:exterior_force_dipole_farfield}
}
&
\parbox[c]{90pt}{%
\centering
Force dipole:
}
&
\parbox[c]{55pt}{%
\centering
\adjustimage{height=34pt,valign=m}
{IllustrationB.pdf}\\[5pt]
}
\\
&
\parbox[c]{175pt}{%
  \centering
  Rotlet:\\[2pt]
  $\displaystyle
  \bU_\drop
=
-\frac{\btau\times\re_0}{8\pi r_0^2}=O(r_0^{-2})$
  Eq.~\eqref{eq:exterior_rotlet_velocity}
}
&
\parbox[c]{100pt}{%
  \centering
  Rotlet:
}
&
\adjustimage{height=50pt,valign=m}
  {IllustrationC.pdf}
\\
\midrule
\parbox[c]{130pt}{%
\centering
General Second Force Moment\\
Sec.~\ref{subsubsec:exteriorquadrupole}\\
$\rff_i^{+}
 =\rM_{ijk}\,\partial_j\partial_k
 \delta(\mathbf x-\mathbf x_0)$
}
&
\parbox[c]{185pt}{%
\centering
$\displaystyle
U_{d,m}
\sim
\frac{\rM_{ijk}}{8\pi}
\left.
\partial_j\partial_k
\right.
$
\\[-1pt]
$\displaystyle
\left.
\times
\left[
\frac{1}{r}
\left(
\delta_{mi}+\frac{x_mx_i}{r^2}
\right)
\right]
\right|_{\mathbf x=\mathbf x_0}
$
\\[-1pt]
$\displaystyle
=O(r_0^{-3})
$ Eq.~\eqref{eq:exterior_force_quadrupole_farfield}
}
&
\parbox[c]{90pt}{%
\centering
Axisymmetric\\
second force moment:
}
&
\adjustimage{height=50pt,valign=m}
{IllustrationD.pdf}
\\
\midrule
\parbox[c]{130pt}{%
\centering
Source Dipole\\
Sec.~\ref{subsubsec:SD}\\
$s^{+}
 =-\mathbf Q\cdot\boldsymbol{\nabla}
 \delta(\mathbf x-\mathbf x_0)$
}
&
\parbox[c]{185pt}{%
\centering
$\displaystyle
\mathbf U_d
=
-\frac{1}{4\pi r_0^3}
\left(
\mathbf I
-3{\mathbf e}_0{\mathbf e}_0
\right)\cdot\mathbf Q
$
\\[2pt]
$\displaystyle
=O(r_0^{-3})
$ Eq.~\eqref{eq:exterior_source_dipole_velocity}
}
&
\parbox[c]{90pt}{%
\centering
Source Dipole:
}
&
\adjustimage{height=50pt,valign=m}
{IllustrationE.pdf}
\\
\midrule
\parbox[c]{130pt}{%
\centering
Higher-order force and source multipoles
\\ Sec.~\ref{subsubsec:HO}}
&
\parbox[c]{185pt}{%
\centering
Rank-$n$ force multipole:
\\[2pt]
$\displaystyle
\mathbf U_d=O(r_0^{-n})
$ Eq.~\eqref{eq:exterior_force_multipole_scaling}
\\[6pt]
Order-$n$ source multipole:
\\[2pt]
$\displaystyle
\mathbf U_d=O(r_0^{-(n+2)})
$ Eq.~\eqref{eq:exterior_source_multipole_scaling}
}
&
&
\\
\bottomrule
\end{tabular}
\end{center}

\caption{Leading far-field contribution to the drop velocity induced by
an exterior singularity, where $r_0=|\mathbf x_0|\gg1$ and
${\mathbf e}_0=\mathbf x_0/r_0$.  The exact expressions in
Sec.~\ref{subsec:exterior_singularities} apply for all $r_0>1$.  Unlike the interior problem, exterior force and source multipoles of arbitrarily high order can contribute to drop translation.}
\label{table:exterior_singularity_summary}
\end{table}

Table~\ref{table:exterior_singularity_summary} also makes the contrast with
Table~\ref{table:singularitysummary} explicit.  For an enclosed singularity,
the polynomial interior auxiliary fields produce a finite selection rule:
force moments above quadrupolar order and source moments above dipolar order
do not contribute directly to translation.  For an exterior singularity, the
nonpolynomial auxiliary fields have nonzero derivatives of every order, so no
such truncation occurs.  In addition, every exterior contribution decays as
the singularity is moved away from the drop.  The leading force-multipole
terms in Table~\ref{table:exterior_singularity_summary} are independent of
$\lambda$ because the factor $3\lambda+2$ in the $O(r^{-1})$ part of
$\bK^\out$ cancels the same factor in Eq.~\eqref{eqRTC:Udropeqn}; viscosity contrast first appears two powers of $r_0$ later.  We note that such dependence on the viscosity ratio $\lambda$ concerns the translational projection only, and does not imply that the complete exterior flow is independent of $\lambda$.

\subsection{Exterior-flow signatures of enclosed point singularities}
\label{subsec:interior_exterior_flow}
Having compared the translational responses to interior and exterior point singularities, we now return to enclosed singularities and ask what additional information is carried by the exterior flow.
The reciprocal relation determines the drop velocity without requiring the
complete flow.  To distinguish enclosed singularities that produce the same
$\bU_\drop$, we also examine their exterior far fields.  Let
$r=|\bx|>1$, $r_0=|\bxz|<1$, $\widehat{\br}=\bx/r$, and, for $r_0>0$,
$\mathbf e_0=\bxz/r_0$.  The translation of spherical Stokes modes from
$\bxz$ to the drop center and their transmission through a clean spherical
interface are standard \cite{felderhof1989displacement,sprenger2020towards,kree2021dynamics,kawakami2025migration}.
We therefore do not repeat the general derivation here. Salient details of the derivations can be found in Appendix~\ref{app:interior_exterior_modes}, where we summarize the notation, transfer
relations, and the low-order calculations.  Here we present the
resulting fields and their small-eccentricity expansion in $r_0 \ll1$. 
The exterior flow fields, written as centered contributions plus their first eccentric corrections, are:
\begin{subequations}
\begin{align}
 \text{Stokeslet:}\quad
 \bu^\out
 &=\frac{(\bI+\widehat{\br}\widehat{\br})\cdot\bF}{8\pi r} +\frac{\lambda-1}{8\pi(3\lambda+2)}\frac{(\bI-3\widehat{\br}\widehat{\br})\cdot\bF}{r^3}+
    \mathcal B_{\rm St}\frac{r_0}{r^2}
   +\cdots,\\
 \text{force dipole:}\quad
 \bu^\out
 &=\frac{\mathcal A_D(\widehat{\br};\bS,\lambda)}{r^2}
   +
    \mathcal A_M(\widehat{\br};\bM^{(D)} ,\lambda)\frac{r_0}{r^3}
   +\cdots ,\\
 \text{general second force moment:}\quad
 \bu^\out
 &=\frac{\mathcal A_M(\widehat{\br};\bM,\lambda)}{r^3}
   +\mathcal B_M(\widehat{\br};\bM,
           \mathbf e_0,\lambda)\frac{r_0}{r^2}+\cdots ,
           \label{eq:ext_flow_2ndfm}\\
 \text{source dipole:}\quad
 \bu^\out
 &=\frac{\mathcal A_{\rm SD}(\widehat{\br};\bQ,\lambda)}{r^3}
   +\mathcal B_{\rm SD}(\widehat{\br};\bQ,
           \mathbf e_0,\lambda)\frac{r_0}{r^2}+\cdots.
           \label{eq:ext_flow_sd}
\end{align}
\label{eq:interior_singularity_exterior_flow_summary}
\end{subequations}

\noindent
The amplitudes $\mathcal B_{\rm St}$,
$\mathcal A_D(\widehat{\br};\bS,\lambda)$,
$\mathcal A_M(\widehat{\br};\bM^{(D)},\lambda)$,
$\mathcal A_M(\widehat{\br};\bM,\lambda)$, 
$\mathcal B_M(\widehat{\br};\bM,
           \mathbf e_0,\lambda)$,
$\mathcal A_{\rm SD}(\widehat{\br};\bQ,\lambda)$ and 
$\mathcal B_{\rm SD}(\widehat{\br};\bQ,
           \mathbf e_0,\lambda)$ 
in the above equations are independent of the offset of the singularity from the drop center, $r_0$, and can be found in Appendix~\ref{app:interior_exterior_modes}.
In the force-dipole expansion, $\bM^{(D)}$ is the second force moment generated at first order when the dipole is displaced from the drop center:
\begin{equation}
\rM_{ijk}^{(D)}
:=
-\rS_{i(j}(e_0)_{k)}
=
-\frac{1}{2}
\left[
\rS_{ij}(e_0)_k
+
\rS_{ik}(e_0)_j
\right].
\label{eq:dipole_induced_second_moment}
\end{equation}
See Appendix~\ref{app:interior_exterior_modes} for the detailed derivation.

\begin{figure}
\includegraphics[width=0.95\linewidth]{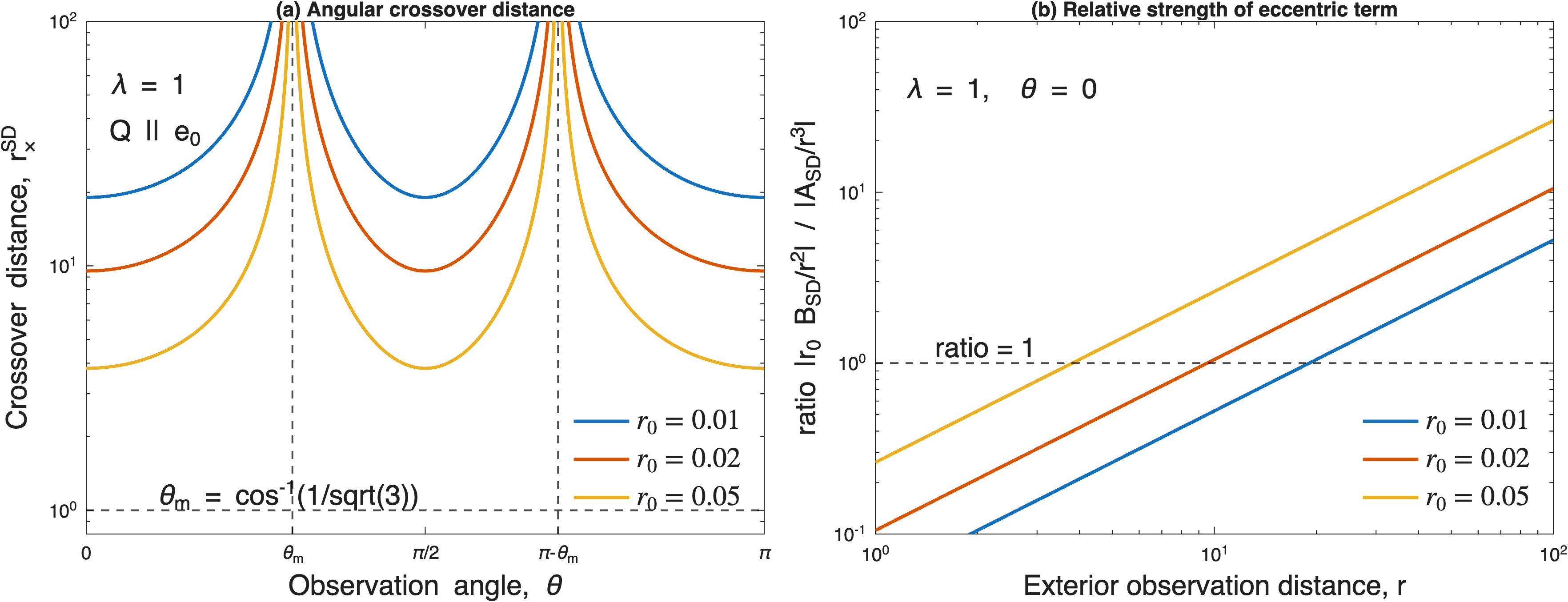}
    \caption{Exterior-flow crossover for an enclosed source dipole with $\lambda=1$ and $\bQ\parallel{\mathbf e}_0$, using the leading small-eccentricity approximation. (a) Crossover distance $r_{\times}^{\rm SD}$ versus observation angle $\theta$ for $r_0=0.01$, $0.02$, and $0.05$ (all smaller than $4/21$). The vertical dashed lines mark $3\cos^2\theta=1$, where the displacement-induced $r^{-2}$ coefficient vanishes and $r_{\times}^{\rm SD}$ diverges; the horizontal dashed line denotes the drop surface $r=1$. (b) Ratio of the displacement-induced $O(r_0r^{-2})$ contribution to the centered $O(r^{-3})$ contribution along $\theta=0$. The crossover occurs where this ratio equals unity.}
    \label{fig:crossover_SD}
\end{figure}

For the exterior flow in Eq.~\eqref{eq:ext_flow_2ndfm} (for a general second force moment) and Eq.~\eqref{eq:ext_flow_sd} (for a source dipole), there is a crossover distance, defined along a fixed observation direction, where the leading order scaling with $r$ changes from $r^{-3}$ to $r^{-2}$.
%
%
Specifically, for the source dipole, the
two displayed terms are equal in magnitude at
\begin{equation}
 r_\times^{\rm SD}(\widehat{\br})
 \sim\frac{10(\lambda+1)}{21(3\lambda+2)r_0}
 \frac{\left|(\bI-3\widehat{\br}\widehat{\br})\cdot\bQ\right|}
 {\left|3(\bQ\cdot\widehat{\br})
          (\mathbf e_0\cdot\widehat{\br})
        -\bQ\cdot\mathbf e_0\right|}, \quad r_0 \ll 1, \label{eq:source_dipole_crossover_distance}
\end{equation}
provided the denominator is nonzero and $r_\times^{\rm SD}>1$.  If
$\bQ\parallel\mathbf e_0$ and $\theta$ is the angle between
$\widehat{\br}$ and $\mathbf e_0$, this reduces to
\begin{equation}
 r_\times^{\rm SD}(\theta)
 \sim\frac{10(\lambda+1)}{21(3\lambda+2)r_0}
 \frac{\sqrt{1+3\cos^2\theta}}
 {|3\cos^2\theta-1|}.
 \label{eq:source_dipole_crossover_axisymmetric}
\end{equation}
Along the symmetry axis ($\theta=0$) and in the equatorial plane
($\theta=\pi/2$), the angular ratio equals one, so
\begin{equation}
 r_\times^{\rm SD}
 \sim \frac{10(\lambda+1)}{21(3\lambda+2)r_0};
 \qquad
 r_\times^{\rm SD}\sim\frac{4}{21r_0},\quad r_0 \ll 1 
 \quad\text{when }\lambda=1.
 \label{eq:source_dipole_crossover_axis_equator}
\end{equation}
For $\lambda=1$ and $\bQ\parallel{\mathbf e}_0$, an exterior crossover along the symmetry axis or equatorial plane requires, to leading order, $r_0 \lesssim 4/21$. At $3\cos^2\theta=1$, the eccentric $r^{-2}$ coefficient vanishes and there is no crossover ($r_\times^{\rm SD}\rightarrow \infty$) at this order, see Fig.~\ref{fig:crossover_SD}(a). For $\theta=0$ the crossover length is expected to decrease with increasing $r_0$, see Fig.~\ref{fig:crossover_SD}(b).

For a general second force moment,
\begin{equation}
 r_\times^{\rM}(\widehat{\br})
 \sim\frac{|\mathcal A_M(\widehat{\br};\rM,\lambda)|}
 {r_0|\mathcal B_M(\widehat{\br};\rM,
          \mathbf e_0,\lambda)|}, \quad r_0 \ll 1, \label{eq:second_moment_crossover_distance}
\end{equation}
when both amplitudes are nonzero.  Thus
$r_\times^{\rM},r_\times^{\rm SD}=O(r_0^{-1})$: at fixed nonzero
$r_0$, the displacement-induced $r^{-2}$ mode eventually exceeds the
centered $r^{-3}$ field.  We discuss the crossover length for a force dipole and a Stokeslet in Appendix~\ref{app:crossover_details}.


Finally, let $\ell=1-r_0$ be the point-to-interface distance.  Note that this near-interface estimate is separate from the small-eccentricity expansion in Eq.~\eqref{eq:interior_singularity_exterior_flow_summary} where $r_0\ll 1$. Focusing on $r_0 \rightarrow 1^\ins$: since
$r_0^j=(1-\ell)^j\simeq e^{-j\ell}$ for $\ell\ll1$, modes up to
$j=O(\ell^{-1})$ are less suppressed (not exponentially small) near the interface.  The near-interface field requires a separate near-field treatment (such as the lubrication scaling in \cite{young2026soft}), whereas the far field remains controlled by the lowest nonzero drop-centered modes. For $r_0 = O(1)$, however, their amplitudes must be evaluated from the full translated-mode representation rather than the small-$r_0$ expansion in Eq.~\eqref{eq:interior_singularity_exterior_flow_summary}.

\section{Regularized distributions and finite-size particles}
\label{sec:finite_size_effects}
We now return to the translational projection and distinguish two notions of finite size for an object contained within the drop. Here we focus on the finite-size effects of a spherical object enclosed inside a spherical drop.
\begingroup\color{black}
A localized object inside the drop can be assigned a finite size in two
fundamentally different ways. First, a point distribution can be replaced by
a smooth blob, a regularized singularity, of ``size" $a_b$; the resulting correction is determined by the
prescribed moments of the regularization kernel. Second, a rigid particle of
radius $a_p$ can be resolved as a fluid-excluding body with a boundary
condition on its surface; the correction is then determined by the traction
produced by the coupled particle--drop problem. The first description
underlies regularized-Stokeslet and force-coupling methods
\cite{Cortez2001,MaxeyPatel2001,CortezFauciMedovikov2005}, whereas the second
represents a physical inclusion. Isotropic second moments generate the
familiar potential- or source-dipole correction
\cite{ZhaoLaugaKoens2019}, and compact immersed-boundary particles and
multiblob methods provide related calibrated representations
\cite{BalboaUsabiagaEtAl2014,DelongEtAl2014,BalboaUsabiagaEtAl2016}.
Although both descriptions enter the same reciprocal identity, their radii
cannot be identified without matching the moments relevant to the drop
velocity.

\subsection{Regularization by a blob of size \texorpdfstring{$a_b$}{ab}}
\label{subsec:finite_blob}

We first define the regularized kernel 
\begin{equation}
\phi_{a_b}(\bx-\bxz)
=
a_b^{-3}\Phi\!\left(\frac{\bx-\bxz}{a_b}\right),
\qquad
\int_{\mathbb R^3}\Phi(\by)\,d\by=1,
\qquad
\int_{\mathbb R^3}\by\Phi(\by)\,d\by=\bzero,
\label{ZeroAndFirstMoments}
\end{equation}
and assume that the kernel is compactly supported and that its support lies
strictly inside $V^\ins$. Its dimensionless second-moment tensor is
\begin{equation}
\rC_{jk}
=
\int_{\mathbb R^3}y_jy_k\Phi(\by)\,d\by.
\label{eq:finite_blob_second_moment}
\end{equation}
For any smooth scalar, vector, or tensor field $g$, a componentwise Taylor expansion about $\bxz$ gives
\begin{equation}
\int_{V^\ins}\phi_{a_b}(\bx-\bxz)g(\bx)\,dV
=
g(\bxz)
+
\frac{a_b^2}{2}\rC_{jk}\partial_j\partial_k g(\bxz)
+
O(a_b^3).
\label{eq:finite_blob_distribution_expansion}
\end{equation}
The first-order term vanishes because the blob is centered. Equivalently,
\begin{equation}
\int_{V^\ins}(\bx-\bxz)_j(\bx-\bxz)_k
\phi_{a_b}(\bx-\bxz)\,dV
=a_b^2\rC_{jk}.
\label{eq:blob_second_moment}
\end{equation}
When $g=\bK^\ins$ in \refeqn{eq:finite_blob_distribution_expansion}, the expansion is exact because the interior kernel is
quadratic. Thus an enclosed regularized Stokeslet is determined completely by
its total force, center, and second moment. If the support intersects the drop
interface, truncation by $V^\ins$ produces boundary terms and the following
moment formulas no longer apply without modification.

For the prescribed regularized force density
\begin{equation}
\bff^\ins(\bx)=\bF\phi_{a_b}(\bx-\bxz),
\label{eq:regularized_stokeslet_definition}
\end{equation}
contraction with Eq.~\eqref{eq:gradgradKminus} gives
\begin{equation}
\bU_\drop^{\mathrm{blob}}
=
\frac{1}{4\pi(3\lambda+2)}
\left[
\bK^\ins(\bxz)
+
a_b^2\left(\bC-2\operatorname{tr}(\bC)\bI\right)
\right]\cdot\bF.
\label{eq:finite_blob_force}
\end{equation}
An anisotropic kernel can therefore change both the magnitude and direction
of the correction. For an isotropic kernel, $\bC=c_2\bI$, and
\begin{equation}
\bU_\drop^{\mathrm{blob}}
=
\frac{1}{4\pi(3\lambda+2)}
\left[
\bK^\ins(\bxz)-5c_2a_b^2\bI
\right]\cdot\bF,
\label{eq:finite_blob_force_isotropic}
\end{equation}
where
$c_2=\frac13\int_{\mathbb R^3}|\by|^2\Phi(\by)\,d\by.$
A uniform volume kernel on the unit ball has $c_2=1/5$, whereas a uniform
shell has $c_2=1/3$. The $a_b^2$ coefficient is therefore fixed by the kernel
shape, not by its nominal width alone.

At fixed multipole strength, replace the point distribution
$\delta(\bx-\bxz)$ by the same centered blob while holding $\rS_{ij}$,
$\rM_{ijk}$, and $\bQ$ fixed. Because the support lies strictly inside
$V^\ins$, integrations by parts generate no interfacial boundary terms. For a
regularized force dipole,
$\rff_i^\ins=\rS_{ij}\partial_j\phi_{a_b}(\bx-\bxz),$
so that Eq.~\eqref{eqRTC:Udropeqn} gives
\begin{align}
U_{\drop,m}
&=
\frac{\rS_{ij}}{4\pi(3\lambda+2)}
\int_{V^\ins}\rK^\ins_{mi}(\bx)
\partial_j\phi_{a_b}(\bx-\bxz)\,dV
= -\frac{\rS_{ij}\partial_j\rK^\ins_{mi}(\bxz)}
{4\pi(3\lambda+2)}.
\label{eq:finite_blob_force_dipole}
\end{align}
Eq.~\eqref{eq:finite_blob_force_dipole} is exactly the point-dipole result in Eq.~\eqref{eq:drop_velocity_force_dipole}.

For a regularized force quadrupole,
$\rff^\ins_i
=\rM_{ijk}\partial_j\partial_k\phi_{a_b}(\bx-\bxz)$,
two integrations by parts give
\begin{align}
U_{\drop,m}
&=
\frac{\rM_{ijk}}{4\pi(3\lambda+2)}
\int_{V^\ins}\phi_{a_b}(\bx-\bxz)
\partial_j\partial_k \rK^\ins_{mi}(\bx)\,dV =
\frac{\rM_{ijk}\partial_j\partial_k K^\ins_{mi}}
{4\pi(3\lambda+2)}.
\label{eq:finite_blob_force_quadrupole}
\end{align}
The second derivative is constant, so normalization alone recovers the point
quadrupole result in Eq.~\eqref{eq:drop_velocity_force_quad}.

Likewise, for the regularized source dipole, $s^\ins=-\rQ_j\partial_j\phi_{a_b}(\bx-\bxz)$, the drop velocity is identical to the result in Eq.~\eqref{eqRTC:RTSD3}:
\begin{align}
U_{\drop,m}
&=
-\frac{5\lambda Q_j}{2\pi(3\lambda+2)}
\int_{V^\ins}\rx_m\partial_j\phi_{a_b}(\bx-\bxz)\,dV
=
\frac{5\lambda \rQ_j}{2\pi(3\lambda+2)}
\int_{V^\ins}\delta_{mj}\phi_{a_b}(\bx-\bxz)\,dV =
\frac{5\lambda}{2\pi(3\lambda+2)}\rQ_m.
\label{eq:finite_blob_source_dipole}
\end{align}
Hence regularization leaves the force-dipole, force-quadrupole, and
source-dipole velocities unchanged at fixed multipole strength. The
Stokeslet is different because it samples $\bK^\ins$ itself, whose nonzero
second derivative couples to the blob's second moment. 

\subsection{A finite-size rigid particle}
\label{subsec:finite_rigid_particle}

Let a rigid spherical particle of radius $a_p$ located at $\bxp$ occupy the volume $B_{a_p}(\bxp)$, with surface
$S_p$. We assume that the entire particle, including its surface, lies in the interior fluid.
Let $\bn_p$ point from the particle into the fluid. The traction exerted by
the particle on the fluid and its resultant force are
$\bg=-\bsigma^\ins\cdot\bn_p,$
$\bF=\int_{S_p}\bg\,dS.$
Applying the reciprocal theorem in
$V^\ins\setminus B_{a_p}(\bxp)$ introduces an integral over $S_p$. For a
no-slip particle, terms proportional to its rigid translation and rotation
vanish because the smooth auxiliary solution satisfies
\begin{equation}
\int_{S_p}\widehat{\bsigma}^\ins\cdot\bn_p\,dS=\bzero,
\qquad
\int_{S_p}(\bx-\bxp)\times
(\widehat{\bsigma}^\ins\cdot\bn_p)\,dS=\bzero.
\label{eq:auxiliary_zero_force_torque}
\end{equation}
The resulting drop velocity is
\begin{equation}
\bU_\drop
=
\frac{1}{4\pi(3\lambda+2)}
\int_{S_p}\bK^\ins(\bx)\cdot\bg(\bx)\,dS.
\label{eq:finite_exact_traction}
\end{equation}
Eq.~\eqref{eq:finite_exact_traction} is exact for a fully enclosed no-slip particle in the spherical-drop problem and does not require $a_p \ll 1$. The radius enters through the integration surface and through the traction $\bg$, which depends on particle position, imposed motion, applied force and torque, surface slip, and hydrodynamic interaction with the drop interface.

If a tangential slip $\bu_s$ is prescribed on $S_p$, its contribution to $\bU_\drop$ can be expressed in the following integral for any unit vector $\widetilde{\bU}$:
\begin{subequations}
\begin{align}
\bU_\drop
 &=
\frac{1}{4\pi(3\lambda+2)}
\int_{S_p}\bK^\ins(\bx)\cdot\bg(\bx)\,dS + \bU_\drop^{(s)},
\label{eq:finite_exact_traction_with_squirmer} \\
\widetilde\bU\cdot\bU_\drop^{(s)}
&=
\frac{\lambda+1}{2\pi(3\lambda+2)}
\int_{S_p}\bu_s\cdot
\left(
\widehat{\bsigma}^\ins[\widetilde{\bU}]\cdot\bn_p
\right)dS.
\label{eq:finite_slip_term}
\end{align}
\end{subequations}
Evaluating Eq.~\eqref{eq:finite_slip_term} for three independent choices of unit vector $\widetilde{\bU}$ determines the vector $\bU_\drop^{(s)}$. Slip also changes $\bg$, so the traction and explicit slip contributions are coupled outputs of the same boundary-value problem.

Define the corresponding traction moments by
\begin{equation}
\rS_{ij}
=-\int_{S_p}g_i(\bx-\bxp)_j\,dS,
\qquad
\rM_{ijk}
=\frac12\int_{S_p}g_i(\bx-\bxp)_j(\bx-\bxp)_k\,dS,
\qquad
\rM_{ijk}=\rM_{ikj}.
\label{eq:finite_traction_moments}
\end{equation}
Because $\bK^\ins$ is quadratic, its Taylor expansion about $\bxp$ terminates,
and Eq.~\eqref{eq:finite_exact_traction} becomes
\begin{equation}
U_{\drop,m}
=
\frac{1}{4\pi(3\lambda+2)}
\left[
\rK^\ins_{mi}(\bxp)\rrF_i
-
\partial_j\rK^\ins_{mi}(\bxp)\rS_{ij}
+
\partial_j\partial_k\rK^\ins_{mi}\rM_{ijk}
\right].
\label{eq:finite_exact_moments}
\end{equation}
This is the exact finite-size particle counterpart of the interior point-multipole
formulas in Eq.~\eqref{eqRTC:Udropeqn}. No traction moment above quadrupolar order contributes directly to
translation. Higher moments may alter the surrounding flow or enter
indirectly by changing the traction, but the translational projection
itself is complete at quadrupolar order.

Using Eqs.~\eqref{eq:gradKminus} and \eqref{eq:gradgradKminus},
\begin{subequations}
\begin{align}
U_{\drop,m}^{(D)}
&=
\frac{4\rS_{mj}\rx_{{\rm p}j}-\rS_{im}\rx_{{\rm p}i}-\rS_{ii}\rx_{{\rm p}m}}
{4\pi(3\lambda+2)},
\label{eq:finite_explicit_dipole}\\
U_{\drop,m}^{(Q)}
&=
\frac{2\rM_{iim}-4\rM_{mjj}}
{4\pi(3\lambda+2)}.
\label{eq:finite_explicit_quadrupole}
\end{align}
\end{subequations}
For symmetric $\bS$, the dipole numerator reduces to
$3\bS\cdot\bxp-\operatorname{tr}(\bS)\bxp$; for a symmetric,
trace-free stresslet, only $3\bS\cdot\bxp$ remains. The quadrupole affects
translation only through the vector traces $\bM_{iim}$ and $\bM_{mjj}$. Although
the explicit dipole weight is linear in $|\bxp|$ and the explicit quadrupole
weight is independent of $\bxp$, the moments themselves generally depend on
particle position and can vary strongly near the interface.

A small forced no-slip sphere provides a limit in which the leading traction
moments can be evaluated explicitly. 
First we define $\ell:=1-|\bxp|,$  and assume $a_p/\ell\ll1$,
so the particle is small compared with its center-to-interface distance. To
leading order, the surrounding flow is the unbounded-fluid solution and the
traction exerted on the fluid is uniform: $\bg=\bF/(4\pi a_p^2)$.
Using
\begin{equation}
\int_{S_p}(\bx-\bxp)_j\,dS=0,
\qquad
\int_{S_p}(\bx-\bxp)_j(\bx-\bxp)_k\,dS
=\frac{4\pi a_p^4}{3}\delta_{jk},
\end{equation}
Eq.~\eqref{eq:finite_traction_moments} gives $\rS_{ij}=0$ and  $\rM_{ijk}=\frac{a_p^2}{6}\rrF_i\delta_{jk}$: The dipole vanishes by spherical symmetry, whereas the isotropic second traction moment generates a general second force moment. Substitution into
Eq.~\eqref{eq:finite_exact_moments} while taking $a_p\to0$ with $\bx_p$ fixed (so that
$\ell=1-|\bx_p|=O(1)$ and $a_p/\ell\ll1$) yields
\begin{equation}
\bU_\drop^{\mathrm{sphere}}
=
\frac{1}{4\pi(3\lambda+2)}
\left[
\left(
2\lambda+3-2|\bxp|^2-\frac53a_p^2
\right)\bI
+
\bxp\bxp
\right]\cdot\bF
+
o(a_p^2|\bF|).
\label{eq:finite_forced_sphere_speed}
\end{equation}
The $a_p^2$ correction (at fixed $\bx_p$) is isotropic and shifts the radial and tangential
response coefficients by the same amount. In the point-multipole
representation, the same result follows from
\begin{equation}
\bff_{\mathrm{eq}}^\ins
=
\bF\delta(\bx-\bxp)
+
\frac{a_p^2}{6}\bF\,\nabla^2\delta(\bx-\bxp).
\label{eq:finite_equivalent_force}
\end{equation}
Indeed, $\nabla^2\bK^\ins=-10\bI$, so the second term contributes
$-(5/3)a_p^2\bF/[4\pi(3\lambda+2)]$.
Equation~\eqref{eq:finite_equivalent_force} is consistent with flow generated by a rigid sphere of radius $a_p$ under a constant force $\bF$ in the unbounded domain \cite{graham2018microhydrodynamics}.

The approximation in Eq.~\eqref{eq:finite_forced_sphere_speed} fails when the
surface gap $h:=1-|\bxp|-a_p$
is comparable with $a_p$. In that regime the interface produces order-one
variations in $\bg$, and the uniform-traction approximation is no longer
valid. The exact identity in Eq.~\eqref{eq:finite_exact_moments} remains valid as long
as the drop is spherical and the actual traction moments are supplied.

Finally, matching the forced-sphere correction to the isotropic blob result
requires
\begin{equation}
c_2a_b^2=\frac{a_p^2}{3}.
\label{eq:finite_blob_particle_match}
\end{equation}
A uniform shell kernel therefore matches when $a_b=a_p$, whereas a uniform volume
kernel requires $a_b=\sqrt{5/3}\,a_p$. This calibration matches only the
projected drop velocity in the small forced-sphere limit; it does not make the
blob and physical particle equivalent at the level of traction or surrounding
flow.
\endgroup

\section{Drop with variable interfacial tension}
\label{sec:variable_tension}

\begingroup\color{black}
The preceding section extends the localized-forcing description to finite-size distributions and rigid particles, while retaining a spherical drop with uniform interfacial tension. A complementary extension is to allow the interface itself to exert a nonuniform stress that may arise from a nonuniform surfactant concentration or as a variable tension enforcing surface inextensibility. We therefore consider a spatially varying interfacial tension $\gamma=\gamma(\theta,\phi)$ and determine its direct contribution to drop translation.  With the traction-jump convention in Eq.~\eqref{eqRTC:BC2},
$\bt_\surf
=
\gamma\kappa\bn-\bnab_\surf\gamma,$
with 
$\kappa=2.$
Using Eq.~\eqref{eqRTC:bushat}, the first surface term in the reciprocal
identity becomes
\begin{equation}
-
\int_\surf \bt_\surf\cdot\widehat{\bu}_\surf\,dS
=
-
\widehat{\bU}\cdot
\int_\surf
\left[
2\gamma\bn
-
\frac{2\lambda+1}{2(\lambda+1)}\bnab_\surf\gamma
\right]dS.
\label{eq:variable_tension_reciprocal_term}
\end{equation}
Let $\bU_\drop^{\mathrm{vol}}$ denote the force-and-source contribution given
by the right-hand side of Eq.~\eqref{eqRTC:Udropeqn}. The total velocity is
then
\begin{equation}
\bU_\drop
=
\bU_\drop^{\mathrm{vol}}
-
\frac{1}{4\pi(3\lambda+2)}
\int_\surf
\left[
4(\lambda+1)\gamma\bn
-
(2\lambda+1)\bnab_\surf\gamma
\right]dS.
\label{eqRTC:Udropeqn2}
\end{equation}
The normal vector multiplying $\gamma$ in Eq.~\eqref{eqRTC:Udropeqn2} is
essential: both terms in the integrand are vectors. On the unit sphere $\int_\surf\bnab_\surf\gamma\,dS = 2\int_\surf\gamma\bn\,dS$
so Eq.~\eqref{eqRTC:Udropeqn2} simplifies to
\begin{equation}
\bU_\drop
=
\bU_\drop^{\mathrm{vol}}
-
\frac{1}{2\pi(3\lambda+2)}
\int_\surf\gamma\bn\,dS.
\label{eq:variable_tension_compact}
\end{equation}
This form immediately shows that a spatially uniform tension produces no
translation. 
Using the mobility relation for a spherical drop, the correction to the drop velocity in Eq.~\ref{eq:variable_tension_compact} can be expressed in terms of a force exerted on a stationary drop:
\begin{flalign}
    \bU_\drop =\bU_\drop^{\mathrm{vol}}+\frac{\viscratio+1}{2\pi(3\viscratio+2)}\bF_\drop, \quad \bF_\drop = -\frac{1}{2(\viscratio+1)}\int_\surf\bnab_\surf \gamma \,d\surf.\label{eq:variable_tension_force}
\end{flalign}
The form for the force in Eq.~\ref{eq:variable_tension_force} matches the force derived originally in Ref. \cite{subramanian1985stokes}.
Expand the tension in the complex spherical-harmonic basis used in Ref.~\cite{kawakami2025migration}:
$\gamma(\theta,\phi)
=
\gamma_0+
\sum_{j=1}^{\infty}\sum_{m=-j}^{j}
\gamma_{j,m}Y_{j,m}(\theta,\phi).$
Only the $j=1$ coefficients survive the integral in
Eq.~\eqref{eq:variable_tension_compact}. Consequently,
\begin{align}
\bU_\drop
&=
\bU_\drop^{\mathrm{vol}}
-
\frac{1}{\sqrt{6\pi}(3\lambda+2)}
\left[
(\gamma_{1,-1}-\gamma_{1,1})\xhat
-i(\gamma_{1,-1}+\gamma_{1,1})\yhat
+\sqrt{2}\gamma_{1,0}\zhat
\right].
\label{eqRTC:Udropeqn3}
\end{align}
Thus the degree-one  ($j=1$) tension mode is the only mode that contributes directly
to translation of an exactly spherical drop. Higher-degree modes can alter
interfacial circulation and stress but have zero direct projection onto the
translating auxiliary mode.
\endgroup

\section{Conclusions and Discussion}

We have derived a reciprocal-theorem representation for the translational
velocity of a spherical viscous drop driven by localized force and source
distributions in either fluid. Equation~\eqref{eqRTC:Udropeqn} projects the
prescribed forcing onto the Hadamard--Rybczynski translating mode and applies
to point singularities, distributed forces, finite-particle tractions, and
interfacial stresses. It gives the instantaneous drop velocity for a prescribed forcing configuration; determining a mobile particle’s translation and rotation requires a separate mobility calculation or force–torque balance coupled to the drop motion.

For localized forcing strictly inside the drop, the quadratic auxiliary velocity
selects only the net force, force dipole, and general second force moment,
while the linear auxiliary pressure selects only the first source moment.
Higher multipoles remain in the physical flow but do not contribute directly
to translation, as summarized in
Table~\ref{table:singularitysummary}. Zero net force therefore removes only
the Stokeslet contribution: an off-center stresslet or rotlet, a source
dipole, the vector-trace part of a general second force moment, or prescribed
slip can still translate the drop. For exterior forcing, the auxiliary
kernel has nonzero derivatives at every order, so no finite multipole
truncation occurs
(Table~\ref{table:exterior_singularity_summary}). For $r_0\gg1$, a rank-$n$
force multipole gives $\bU_\drop=O(r_0^{-n})$, whereas an order-$n$ source
multipole gives $O(r_0^{-(n+2)})$. The leading force-multipole response is
independent of $\lambda$; viscosity-dependent corrections generally enter
two powers of $r_0$ later.

Distinct enclosed forcing distributions can have the same projected moments
and hence the same $\bU_\drop$. Their exterior flows nevertheless retain
additional information. Expansion in drop-centered spherical Stokes modes
shows, for example, that displacing a Stokeslet changes the exterior flow at
$O(r_0r^{-2})$ although its drop velocity changes only at $O(r_0^2)$, while
a displaced source dipole excites a stresslet-like $O(r_0r^{-2})$ mode
without changing $\bU_\drop$. These signatures constrain singularity type,
orientation, and position beyond what can be inferred from translation
alone. As the point-to-interface distance $\ell=1-r_0$ decreases, modes up
to $j=O(\ell^{-1})$ become relevant, indicating the increasing complexity of
the near-interface flow.

The finite-size analysis distinguishes a regularized distribution from a
resolved particle. A prescribed blob introduces finite size through its
kernel moments: a regularized Stokeslet samples the second kernel moment,
whereas fixed-strength force dipoles, general second force moments, and
source dipoles retain their point-singularity velocities. A rigid particle
instead excludes fluid and generates a traction determined by the coupled
particle--drop problem. Equation~\eqref{eq:finite_exact_moments} reduces the
no-slip traction contribution exactly to the net force and first two
traction moments, while prescribed slip adds a separate contribution.
Particle size and position alone therefore do not determine the correction
because loading, torque, slip, and interface interaction also affect these
moments.

For a small forced no-slip sphere far from the interface, symmetry removes
the dipole moment and the first correction is proportional to $a_p^2$.
Matching this correction to an isotropic blob calibrates the blob size only
for the projected drop velocity, not for the traction or exterior flow. As
the particle--interface gap narrows, the exact moment identity remains valid,
but the gap-scale traction must be resolved. For prescribed tension on an
exactly spherical interface, only the degree-one spherical-harmonic
component contributes directly to translation.

These results provide benchmarks for numerical and experimental models.
Regularized-singularity formulas and small-particle asymptotics test
particle--drop simulations, while
Eq.~\eqref{eq:finite_exact_moments} compares a computed drop velocity with
low-order moments of the resolved traction. Such checks are especially useful
near contact, where layer-potential evaluations become nearly singular
\cite{TlupovaBeale2013,afKlintebergTornberg2016,young2026soft}.
Measurements of $\bU_\drop(t)$ and low-order exterior-flow modes obtained by
particle image velocimetry or particle tracking
\cite{kokot2022spontaneous,VincentiEtAl2019} can be compared with candidate
singularity models \cite{kawakami2025migration}. Together, these observables can constrain the position, orientation, and admissible multipole content of the enclosed activity. The same framework can also help guide the design of squirmer-like viscous drops (ongoing work) for controlled caged swimming via reinforcement learning \cite{zou2022gait,zou2024adaptive}.

\begin{acknowledgments}
S. K. and Y. N. Y. acknowledge support
from NSF (DMS-1951600 and DMS-2510714). Y. N. Y.
also acknowledges the Flatiron Institute, part of Simons Foundation. 
H. A. S. acknowledges support from NSF (CBET-224679).
\end{acknowledgments}

\appendix
\section{Exterior modes generated by an enclosed singularity}
\label{app:interior_exterior_modes}

This appendix collects the spherical-mode formulas used in
Sec.~\ref{subsec:interior_exterior_flow}.  The displacement theorem for
spherical Stokes solutions is due to Felderhof and Jones
\cite{felderhof1989displacement}.  Its implementation in the present
vector-spherical-harmonic basis, together with the clean-drop transmission
relations and explicit Stokeslet, rotlet, and stresslet transforms, is given
in Ref.~\cite{kawakami2025migration}, especially Sec.~3 and Appendices A and
B.  We use these established results and retain only the low-order formulas
needed for the eccentricity expansions in the main text.

Throughout this appendix, $\bS$ and $\bM$ denote the force-dipole and general
second-force-moment tensors as objects, whereas $\rS_{ij}$ and $\rM_{ijk}$
denote their components.  We write $r=|\bx|$, $r_0=|\bxz|$,
$\widehat{\br}=\bx/r$, and $\mathbf e_0=\bxz/r_0$ for $r_0>0$.

\subsection{Drop-centered modes and clean-interface transmission}
\label{app:mode_transfer}

Define the vector spherical harmonics
\begin{equation}
 \mathbf y_{jm0}=\frac{\nabla_sY_{jm}}{\sqrt{j(j+1)}},\qquad
 \mathbf y_{jm1}=-i\widehat{\br}\times\mathbf y_{jm0},\qquad
 \mathbf y_{jm2}=Y_{jm}\widehat{\br}.
 \label{eq:vector_spherical_harmonics_for_projection}
\end{equation}
To avoid confusion with the superscripts $\ins$ and $\out$ used for the
interior and exterior fluids, we denote the decaying spherical Stokes basis
by $\bu^{\rm dec}_{jm\sigma}$.  In the notation of
Ref.~\cite{kawakami2025migration}, these fields are
$\bu^-_{jm\sigma}$ and are
\begin{subequations}
\label{eq:decaying_spherical_stokes_basis}
\begin{align}
 \bu^{\rm dec}_{jm0}
 &=\frac12 r^{-j}(2-j+jr^{-2})\mathbf y_{jm0}
 +\frac12 r^{-j}\sqrt{j(j+1)}(1-r^{-2})\mathbf y_{jm2},\\
 \bu^{\rm dec}_{jm1}&=r^{-j-1}\mathbf y_{jm1},\\
 \bu^{\rm dec}_{jm2}
 &=\frac12 r^{-j}(2-j)\sqrt{\frac{j}{j+1}}
   (1-r^{-2})\mathbf y_{jm0}
 +\frac12 r^{-j}[j+(2-j)r^{-2}]\mathbf y_{jm2}.
\end{align}
\end{subequations}
They are normalized by
$\bu^{\rm dec}_{jm\sigma}|_{r=1}=\mathbf y_{jm\sigma}$.
Hence, for $r>r_0$, the free-space field generated by a singularity at
$\bxz$ can be expanded as
\begin{subequations}
\begin{align}
 \bu^{\rm sing}(\bx;\bxz)
 &=\sum_{j=1}^{\infty}\sum_{m=-j}^{j}\sum_{\sigma=0}^{2}
 a_{jm\sigma}(\bxz)\bu^{\rm dec}_{jm\sigma}(\bx),
 \label{eq:singularity_drop_center_expansion}\\
 a_{jm\sigma}(\bxz)
 &=\int_{S^2}\mathbf y_{jm\sigma}^{*}\cdot
   \bu^{\rm sing}(\widehat{\br};\bxz)\,d\Omega.
 \label{eq:singularity_projection_coeff}
\end{align}
\end{subequations}
The first relation may equivalently be generated by the displacement theorem for the Stokes equation basis \cite{felderhof1989displacement}.
The transmitted exterior field is
\begin{equation}
 \bu^\out(\bx)
 =\sum_{j=1}^{\infty}\sum_{m=-j}^{j}\sum_{\sigma=0}^{2}
 c_{jm\sigma}\bu^{\rm dec}_{jm\sigma}(\bx).
 \label{eq:interior_singularity_exterior_flow_expansion}
\end{equation}
Our $a_{jm\sigma}$ correspond to the drop-centered decaying coefficients
$c_{jm\sigma}^{\mathrm{act},-}$ in
Ref.~\cite{kawakami2025migration}; $c_{jm\sigma}$ include the response of
the clean interface.

For $j=1$, clean-interface matching gives
\begin{equation}
 \begin{pmatrix}c_{1m0}\\[2pt]c_{1m2}\end{pmatrix}
 =\frac{\lambda}{3\lambda+2}
 \begin{pmatrix}
 2\lambda+3 & \sqrt{2}(\lambda-1)\\
 \sqrt{2}(\lambda-1) & \lambda+4
 \end{pmatrix}
 \begin{pmatrix}a_{1m0}\\[2pt]a_{1m2}\end{pmatrix},
 \qquad c_{1m1}=\lambda a_{1m1}.
 \label{eq:interior_to_exterior_transfer_j1}
\end{equation}
For $j\ge2$, the steady spherical clean-drop problem has no shape-changing
normal mode, $c_{jm2}=0$, and
\begin{subequations}
\label{eq:interior_to_exterior_transfer}
\begin{align}
 c_{jm0}
 &=\frac{\lambda}{\lambda+1}
 \left(2a_{jm0}-\frac{3a_{jm2}}{\sqrt{j(j+1)}}\right),\\
 c_{jm1}
 &=\frac{\lambda(2j+1)}
 {2j+1+(\lambda-1)(j-1)}a_{jm1},\\
 c_{jm2}&=0.
\end{align}
\end{subequations}
The individual basis fields satisfy
$\bu^{\rm dec}_{jm0},\bu^{\rm dec}_{jm2}=O(r^{-j})$ and
$\bu^{\rm dec}_{jm1}=O(r^{-j-1})$.  A physical singularity can decay
faster when leading terms in a linear combination cancel.  For example,
$a_{jm2}=-\sqrt{(j+1)/j}\,a_{jm0}$ cancels the nominal $r^{-j}$ polar
contribution.  Because the interface changes the relative modal weights,
such a cancellation need not persist in the transmitted field.

\subsection{Low-order amplitudes used in Sec.~\ref{subsec:interior_exterior_flow}}
\label{app:second_moment_amplitudes}

\subsubsection{Stokeslet and force dipole}

Decompose the force-dipole tensor into its symmetric trace-free
and antisymmetric parts 
 \begin{equation}
  \bE=\frac{\bS+\bS^{\mathsf T}}{2}
       -\frac{\operatorname{tr}(\bS)}{3}\bI,
  \qquad
  (\tau_S)_i=\epsilon_{ijk}\rS_{jk}.
  \label{eq:dipole_irreducible_parts}
 \end{equation}
For the pure rotlet convention used in
Eq.~\eqref{eq:rotlet_force_density}, $\boldsymbol{\tau}_S=\btau$.
The dominant exterior field of a centered force dipole is therefore
\begin{equation}
 \mathcal A_D(\widehat{\br};\bS,\lambda)
 =\frac{9}{16\pi(\lambda+1)}
   (\widehat{\br}\cdot\bE\cdot\widehat{\br})\widehat{\br}
 +\frac{1}{8\pi}\,\boldsymbol{\tau}_S\times\widehat{\br},
 \qquad
 \bu_D^\out=\frac{\mathcal A_D}{r^2}+O(r^{-4}).
 \label{eq:centered_force_dipole_exterior_amplitude}
\end{equation}
In particular,
\begin{equation}
 \bu^\out_{\rm R}(\bx;\bzero)
 =\frac{1}{8\pi}\frac{\btau\times\bx}{r^3}.
 \label{eq:exterior_flow_centered_interior_rotlet}
\end{equation}
For a centered Stokeslet,
\begin{equation}
 \bu^\out_{\rm St}(\bx;\bzero)
 =\frac{1}{8\pi r}
  (\bI+\widehat{\br}\widehat{\br})\cdot\bF
 +\frac{\lambda-1}{8\pi(3\lambda+2)r^3}
  (\bI-3\widehat{\br}\widehat{\br})\cdot\bF.
 \label{eq:exterior_flow_centered_interior_stokeslet}
\end{equation}
The translation
\begin{equation}
 \rrF_i\delta(\bx-\bxz)
 =\rrF_i\delta(\bx)
 -r_0\rrF_i(e_0)_j\partial_j\delta(\bx)+O(r_0^2)
 \label{eq:stokeslet_distribution_translation}
\end{equation}
shows that its $O(r_0)$ correction is the centered force dipole
$\rS_{ij}=-\rrF_i(e_0)_j$.  Substitution into
Eq.~\eqref{eq:centered_force_dipole_exterior_amplitude} gives
\begin{equation}
 \mathcal B_{\rm St}(\widehat{\br};\bF,\mathbf e_0,\lambda)
 =\frac{3}{16\pi(\lambda+1)}
 \left[\bF\cdot\mathbf e_0
 -3(\bF\cdot\widehat{\br})
    (\mathbf e_0\cdot\widehat{\br})\right]\widehat{\br}
 +\frac{1}{8\pi}
 \left[(\mathbf e_0\times\bF)\times\widehat{\br}\right].
 \label{eq:stokeslet_first_eccentric_amplitude}
\end{equation}
Thus eccentricity first appears in the exterior flow at
$r_0\mathcal B_{\rm St}/r^2$, although the drop velocity in
Eq.~\eqref{eqRTC:RTStokeslet1} changes only at $O(r_0^2)$.

A displaced force dipole similarly generates, at first order, the second
force moment
\begin{equation}
 \rM^{(D)}_{ijk}
 :=-\rS_{i(j}(e_0)_{k)}
 =-\frac12\left[\rS_{ij}(e_0)_k+\rS_{ik}(e_0)_j\right].
 \label{eq:dipole_induced_second_moment}
\end{equation}
The actual displacement-induced moment is $r_0\bM^{(D)}$.  Hence
\begin{equation}
 \bu_D^\out(\bx;\bxz)
 =\frac{\mathcal A_D(\widehat{\br};\bS,\lambda)}{r^2}
 +\frac{r_0}{r^3}
  \mathcal A_M(\widehat{\br};\bM^{(D)},\lambda)
 +\cdots .
 \label{eq:force_dipole_exterior_small_eccentricity}
\end{equation}
The amplitude $\mathcal A_M$ is derived next.

\subsubsection{General second force moment}

Let $\rM_{ijk}=\rM_{ikj}$.  To decompose the centered field, define the two
vector traces
$({\rm m}^{(1)})_i=\rM_{ijj}$,
$(m^{(2)})_i=\rM_{jji},$
$\mathbf P=2\mathbf m^{(1)}-\mathbf m^{(2)}.$
For the degree-two part, set
\begin{equation}
 {\rm N}_{pq}=\epsilon_{pij}\rM_{ijq},\qquad
 \mathsf T_{pq}
 =\frac{N_{pq}+N_{qp}}{2}-\frac{\delta_{pq}}{3}N_{kk}.
 \label{eq:second_moment_toroidal_irreducible}
\end{equation}
For the fully symmetric part, define
\begin{subequations}
\label{eq:second_moment_polar_irreducible}
\begin{align}
 \widetilde S_{ijk}
 &=\rM_{(ijk)}
 =\frac{\rM_{ijk}+\rM_{jik}+\rM_{kij}}{3},\\
 \mathsf H_{ijk}
 &=\widetilde S_{ijk}
 -\frac15\left(\delta_{ij}s_k+\delta_{ik}s_j+\delta_{jk}s_i\right),\quad s_i=\widetilde S_{jji}.
\end{align}
\end{subequations}
Thus $\mathbf P$, $\mathsf T$, and $\mathsf H$ are, respectively, the
degree-one potential, degree-two toroidal, and degree-three polar pieces of
the centered second-moment field.

For a direct projection, introduce
$(c_{\rm M})_i=\rM_{ijk}\widehat r_j\widehat r_k,$
 $(d_M)_i=\rM_{jik}\widehat r_j\widehat r_k,$ 
 and $\chi_M=\rM_{ijk}\widehat r_i\widehat r_j\widehat r_k.$
%
Differentiating the Oseen tensor twice gives the free-space velocity on the
unit sphere,
\begin{align}
 \bv_M(\widehat{\br})
 =\frac{1}{8\pi\lambda}\Big[&-\mathbf m^{(1)}+2\mathbf m^{(2)}
 +3\mathbf c_M-6\mathbf d_M
 -3\widehat{\br}(\mathbf m^{(1)}\cdot\widehat{\br}) -6\widehat{\br}(\mathbf m^{(2)}\cdot\widehat{\br})
 +15\chi_M\widehat{\br}\Big].
 \label{eq:centered_second_moment_free_space_boundary_field}
\end{align}
Projection onto the basis in
Eq.~\eqref{eq:decaying_spherical_stokes_basis}, followed by
Eqs.~\eqref{eq:interior_to_exterior_transfer_j1} and
\eqref{eq:interior_to_exterior_transfer}, yields the centered amplitude
\begin{align}
 \mathcal A_M(\widehat{\br};\bM,\lambda)
 ={}&\frac{1}{4\pi(3\lambda+2)}
 (\bI-3\widehat{\br}\widehat{\br})\cdot\mathbf P
 +\frac{5}{2\pi(\lambda+4)}
 \widehat{\br}\times(\mathsf T\cdot\widehat{\br})
 +\frac{15}{16\pi(\lambda+1)}
 \left[
 \mathsf H:\widehat{\br}\widehat{\br}
 -5(\mathsf H\mathbin{\vdots}
      \widehat{\br}\widehat{\br}\widehat{\br})\widehat{\br}
 \right].
 \label{eq:centered_second_moment_exterior_amplitude}
\end{align}
Here
$(\mathsf H:\widehat{\br}\widehat{\br})_i
={\rm H}_{ijk}\widehat r_j\widehat r_k$ and
$\mathsf H\mathbin{\vdots}\widehat{\br}\widehat{\br}\widehat{\br}
={\rm H}_{ijk}\widehat r_i\widehat r_j\widehat r_k$.
This finite decomposition replaces an infinite modal sum for the centered
second force moment.

For the first eccentric correction, define the symmetric trace-free operator
\begin{equation}
 \mathsf{STF}_{pq}[X]
 :=\frac{X_{pq}+X_{qp}}{2}-\frac{\delta_{pq}}{3}X_{kk},
 \label{eq:rank_two_stf_definition}
\end{equation}
and the degree-two tensor
\begin{align}
 \mathsf K_{pq}(\bM,\mathbf e_0)
 :={}&12\,\mathsf{STF}_{pq}\!\left[(e_0)_i\rM_{ipq}\right]
 -60\,\mathsf{STF}_{pq}\!\left[
 \frac{(e_0)_j(\rM_{pjq}+\rM_{qjp})}{2}\right]
 \notag\\
 &-30\,\mathsf{STF}_{pq}\!\left[
 \frac{(e_0)_p(m^{(1)})_q+(e_0)_q(m^{(1)})_p}{2}\right]
 +24\,\mathsf{STF}_{pq}\!\left[
 \frac{(e_0)_p(m^{(2)})_q+(e_0)_q(m^{(2)})_p}{2}\right].
 \label{eq:second_moment_offset_tensor}
\end{align}
Then
\begin{equation}
 \mathcal B_M(\widehat{\br};\bM,\mathbf e_0,\lambda)
 =-\frac{1}{16\pi(\lambda+1)}
 \left[\mathsf K(\bM,\mathbf e_0):
       \widehat{\br}\widehat{\br}\right]\widehat{\br}.
 \label{eq:second_moment_first_eccentric_amplitude}
\end{equation}
To see the origin of this term, expand
\begin{equation}
 \rM_{ijk}\partial_j\partial_k\delta(\bx-\bxz)
 =\rM_{ijk}\partial_j\partial_k\delta(\bx)
 -r_0\rM_{ijk}(e_0)_l
  \partial_j\partial_k\partial_l\delta(\bx)+O(r_0^2).
 \label{eq:second_moment_distribution_translation}
\end{equation}
The degree-two radial part of the $O(r_0)$ free-space field on $r=1$ is
\begin{equation}
 u^{\rm sing}_{r,j=2}
 =\frac{r_0}{56\pi\lambda}
 \left[\mathsf K(\bM,\mathbf e_0):
       \widehat{\br}\widehat{\br}\right].
 \label{eq:second_moment_incident_j2_radial_field}
\end{equation}
For a degree-two potential combination,
$a_{2m2}=-\sqrt{3/2}\,a_{2m0}$.  Equation~\eqref{eq:interior_to_exterior_transfer}
therefore multiplies this incident radial field by
$-7\lambda/[2(\lambda+1)]$, giving
Eq.~\eqref{eq:second_moment_first_eccentric_amplitude}.  Consequently,
\begin{equation}
 \bu_M^\out(\bx;\bxz)
 =\frac{\mathcal A_M(\widehat{\br};\bM,\lambda)}{r^3}
 +\frac{r_0\mathcal B_M(\widehat{\br};\bM,
        \mathbf e_0,\lambda)}{r^2}
 +\cdots .
 \label{eq:second_moment_exterior_small_eccentricity}
\end{equation}
The $r^{-2}$ coefficient may vanish for special tensor symmetries or
observation directions.

\subsubsection{Source dipole}

For the source-dipole convention in
Eq.~\eqref{eq:source_dipole_definition}, the centered exterior amplitude is
\begin{equation}
 \mathcal A_{\rm SD}(\widehat{\br};\bQ,\lambda)
 =-\frac{5\lambda}{4\pi(3\lambda+2)}
   (\bI-3\widehat{\br}\widehat{\br})\cdot\bQ.
 \label{eq:source_dipole_centered_amplitude}
\end{equation}
The centered free-space field is
\begin{equation}
 \bu_{\rm SD}^{\rm sing}(\bx;\bzero)
 =-\frac{1}{4\pi r^3}
 (\bI-3\widehat{\br}\widehat{\br})\cdot\bQ.
 \label{eq:free_space_source_dipole_centered_appendix}
\end{equation}
Expanding
$\bu_{\rm SD}^{\rm sing}(\bx;\bxz)
 =\bu_{\rm SD}^{\rm sing}(\bx;\bzero)
 -r_0(\mathbf e_0\cdot\nabla)
  \bu_{\rm SD}^{\rm sing}(\bx;\bzero)+O(r_0^2)$,
the degree-two radial part on $r=1$ is
\begin{equation}
 u^{\rm sing}_{r,j=2}
 =\frac{3r_0}{4\pi}
 \left[3(\bQ\cdot\widehat{\br})
          (\mathbf e_0\cdot\widehat{\br})
       -\bQ\cdot\mathbf e_0\right].
 \label{eq:source_dipole_incident_j2_radial_field}
\end{equation}
Applying the same degree-two transfer factor
$-7\lambda/[2(\lambda+1)]$ gives
\begin{equation}
 \mathcal B_{\rm SD}(\widehat{\br};\bQ,\mathbf e_0,\lambda)
 =-\frac{21\lambda}{8\pi(\lambda+1)}
 \left[3(\bQ\cdot\widehat{\br})
          (\mathbf e_0\cdot\widehat{\br})
       -\bQ\cdot\mathbf e_0\right]\widehat{\br}.
 \label{eq:source_dipole_first_eccentric_amplitude}
\end{equation}
Hence
\begin{equation}
 \bu_{\rm SD}^\out(\bx;\bxz)
 =\frac{\mathcal A_{\rm SD}(\widehat{\br};\bQ,\lambda)}{r^3}
 +\frac{r_0\mathcal B_{\rm SD}(\widehat{\br};\bQ,
        \mathbf e_0,\lambda)}{r^2}+\cdots .
 \label{eq:source_dipole_exterior_small_eccentricity}
\end{equation}
This is consistent with Eq.~\eqref{eqRTC:RTSD3}: the source-dipole
contribution to $\bU_\drop$ is independent of $r_0$, whereas its exterior
flow changes at first order in $r_0$.
For $\rM_{ijk}=mF_i\delta_{jk}$ and
$\bQ=-m\bF/\lambda$, Eq.~\eqref{eq:second_moment_first_eccentric_amplitude} reduces to
Eq.~\eqref{eq:source_dipole_first_eccentric_amplitude}; the source-dipole result is therefore
contained in the general second-force-moment expression.

\subsection{Crossover distances}
\label{app:crossover_details}

Along a fixed observation direction, suppose two terms have the form
$\mathbf A/r^p+r_0\mathbf B/r^q$.  If $q<p$, the slower
displacement-induced term exceeds the centered term beyond
\begin{equation}
 r_\times
 =\left(\frac{|\mathbf A|}{r_0|\mathbf B|}\right)^{1/(p-q)}.
 \label{eq:generic_crossover_slower_offset}
\end{equation}
If $q>p$, their formal equality occurs at
\begin{equation}
 r_\times
 =\left(\frac{r_0|\mathbf B|}{|\mathbf A|}\right)^{1/(q-p)}.
 \label{eq:generic_crossover_faster_offset}
\end{equation}
Equation~\eqref{eq:source_dipole_crossover_distance} follows by inserting
the amplitudes derived above. The crossover distance $r_\times^{\rm SD}$ for $\lambda = 1$ and $\bQ\parallel{\mathbf e}_0$ is summarized in Fig.~\ref{fig:crossover_SD}. 
For a force dipole, equality of the two displayed
terms would instead occur at
\begin{equation}
 r_\times^{D}(\widehat{\br})
 \sim r_0\frac{|\mathcal A_M(\widehat{\br};\rM^{(D)},\lambda)|}
 {|\mathcal A_D(\widehat{\br};\bS,\lambda)|}.
 \label{eq:force_dipole_formal_crossover}
\end{equation}
This distance is $O(r_0)<1$ and therefore does not lie in the exterior
region for $r_0\ll1$, unless the centered angular amplitude vanishes.

For the Stokeslet, the $r^{-1}$ term always controls the far field.  The two
subleading terms in Eq.~\eqref{eq:interior_singularity_exterior_flow_summary}
become equal at
\begin{equation}
 r_{\times,\rm sub}^{\rm St}(\widehat{\br})
 =\frac{|\lambda-1|}{8\pi(3\lambda+2)r_0}
 \frac{|(\bI-3\widehat{\br}\widehat{\br})\cdot\bF|}
 {|\mathcal B_{\rm St}(\widehat{\br};\bF,\mathbf e_0,\lambda)|},
 \label{eq:stokeslet_subleading_crossover}
\end{equation}
provided $\mathcal B_{\rm St}\neq\bzero$.  This crossover changes only the
ordering of the subleading corrections and never replaces the leading
Stokeslet.  For $\lambda=1$, the centered potential-dipole term vanishes and
$r_0\mathcal B_{\rm St}/r^2$ is the first correction.

\bibliography{refs}

\end{document}